\documentclass[final,onefignum,onetabnum]{siamart190516}
\usepackage{amsmath,amssymb,url}
\usepackage{hyperref}
\usepackage{cleveref}
\usepackage{bm}
\usepackage{color}
\usepackage{framed}
\usepackage[caption=false,font=footnotesize]{subfig}
\usepackage{floatpag}
\usepackage{multirow}
\usepackage{graphics}

\definecolor{royalblue}{rgb}{0.254901960784,0.411764705882,0.882352941176}
\definecolor{cornflowerblue}{rgb}{0.392156862745,0.58431372549,0.929411764706}
\definecolor{aquamarine}{rgb}{0.498039215686,1.0,0.83137254902}
\definecolor{turquoise}{rgb}{0.250980392157,0.878431372549,0.81568627451}
\definecolor{lightseagreen}{rgb}{0.125490196078,0.698039215686,0.666666666667}
\definecolor{lightcyan}{rgb}{0.878431372549,1.0,1.0}
\definecolor{paleturquoise}{rgb}{0.462745098039,0.627450980392,0.627450980392}
\definecolor{aqua}{rgb}{0.0,1.0,1.0}
\definecolor{lightblue}{rgb}{0.678431372549,0.847058823529,0.901960784314}
\definecolor{deepskyblue}{rgb}{0.0,0.749019607843,1.0}
\definecolor{darkorchid}{rgb}{0.6,0.196078431373,0.8}
\definecolor{mediumorchid}{rgb}{0.729411764706,0.333333333333,0.827450980392}
\definecolor{blueviolet}{rgb}{0.541176470588,0.16862745098,0.886274509804}
\definecolor{forestgreen}{rgb}{0.133333333333,0.545098039216,0.133333333333}
\definecolor{green}{rgb}{0.0,0.501960784314,0.0}
\definecolor{darkgreen}{rgb}{0.0,0.392156862745,0.0}
\definecolor{lightcoral}{rgb}{0.941176470588,0.501960784314,0.501960784314}
\definecolor{firebrick}{rgb}{0.698039215686,0.133333333333,0.133333333333}
\definecolor{tomato}{rgb}{1.0,0.388235294118,0.278431372549}
\definecolor{darkorange}{rgb}{1.0,0.549019607843,0.0}
\definecolor{crimson}{rgb}{0.862745098039,0.078431372549,0.235294117647}
\definecolor{violet}{rgb}{0.933333333333,0.509803921569,0.933333333333}
\definecolor{fuchsia}{rgb}{1.0,0.0,1.0}
\definecolor{yellow}{rgb}{1.0,1.0,0.0}

\def\deepskybluebar#1{{\color{deepskyblue}\rule{8pt}{#1cm}}}
\def\bluebar#1{{\color{blue}\rule{8pt}{#1cm}}}
\def\greenbar#1{{\color{green}\rule{8pt}{#1cm}}}

\ifpdf
\hypersetup{
	pdftitle={A Twitter network and discourse analysis of the Rana Plaza collapse},
	pdfauthor={K. Bergermann, and M. Wolter}
}
\fi

\headers{A Twitter analysis of the Rana Plaza collapse}{K. Bergermann, and M. Wolter}

\title{A Twitter network and discourse analysis of the Rana Plaza collapse}

\author{Kai Bergermann\thanks{Faculty of Mathematics, Technische Universit\"at Chemnitz, 09107 Chemnitz, Germany
		(\email{kai.bergermann@math.tu-chemnitz.de}).}
	\and Margitta Wolter\thanks{Institute of Communication and Media Studies, Universit\"at Leipzig, 04081 Leipzig, Germany.}
}

\begin{document}

\maketitle

\begin{abstract}
Ten years after the collapse of the Rana Plaza textile factory in Dhaka, Bangladesh that killed over $1\,000$ factory workers, the event has become a symbol for the desolate working conditions in fast fashion producer countries in the global south.
We analyze the global Twitter discourse on this event over a three week window around the collapse date over the years $2013$ to $2022$ by a mixture of network-theoretic quantitative and discourse-theoretic qualitative methods.
In particular, key communicators and the community structure of the discourse participants are identified using a multilayer network modeling approach and the interpretative patterns of the key communicator's tweets of all years are analyzed using the sociology of knowledge approach to discourse.
This combination of quantitative and qualitative methods reveals that the discourse is separated into three phases: reporting, reprocessing, and commemoration.
These phases can be identified by the temporal evolution, network-structural properties, and the contentual analysis of the discourse.
After the negotiation of the interpretative framework in the reprocessing phase, subsequent years are characterized by its commemorative repetition as well as resulting demands by different international actor groups despite highly fluctuating participants.
\end{abstract}


\section{Introduction}\label{sec:intro}

The Rana Plaza factory collapse on April $24$th $2013$ triggered major international repercussions including a large body of scholarly works on topics such as worker's safety in the textile industry, international corporate law, or social movements for their improvement \cite{reinecke2015after,siddiqui2016human,barua2017workplace,chowdhury2017rana,bair2020political,rahman2020multi}.
The online discourse on the Rana Plaza disaster remains largely unexplored although social media communication has become an important part of the general public discourse.
Recent social media studies focus on the analysis of online social media platforms such as Twitter by employing quantitative and qualitative methods from various disciplines \cite{waters2011tweet,papadopoulos2012community,ch2015local,omodei2015characterizing,pina2016towards,dickison2016multilayer,riquelme2016measuring,yaqub2017analysis,wu2018emotional,hanteer2019innovative,rehman2020identification,sadri2020exploring,wicke2021covid,shea2022david,wiggins2022nothing,reguero2023journalism}.
In this work, we combine quantitative and qualitative methodology to analyze the international Twitter discourse on the Rana Plaza factory collapse.

The quantitative analysis of this paper builds on tools from complex network science.
Complex systems from various disciplines can effectively be modeled by networks recording pairwise interactions of their entities \cite{watts1998collective,newman2003structure}.
In particular, they form the basis for the quantitative analysis of social networks \cite{scott2012social,brandes2013social,borgatti2018analyzing}.
In recent years, multilayer networks that separately record different types of interactions between the same entities by means of different layers have gained increased attention in many fields including the social sciences \cite{kivela2014multilayer,boccaletti2014structure,omodei2015characterizing,dickison2016multilayer}.

Recent decades have witnessed an increased understanding of complex systems by structural and dynamical network measures.
Among the most prominent structural network analysis techniques are centrality measures \cite{freeman1977set,freeman1978centrality,bonacich1987power,brin1998anatomy,kleinberg1999authoritative,page1999pagerank} and community detection methods \cite{girvan2002community,newman2006modularity,von2007tutorial,fortunato2010community}.
In particular, current research efforts focusing on the analysis of social networks such as Twitter heavily rely on the detection of central users \cite{ch2015local,dickison2016multilayer,riquelme2016measuring,sadri2020exploring} and community structure \cite{papadopoulos2012community,dickison2016multilayer,hanteer2019innovative,rehman2020identification}.

Centrality measures identify and rank the most influential entities in complex networks and some of the earliest contributions in the field were made in the context of social networks \cite{katz1953new,freeman1977set,freeman1978centrality}.
These measures rely on the mathematical formulation of a network in terms of an adjacency matrix and well-studied linear algebraic objects such as eigenvectors are capable of revealing substantial insights into complex systems such as the internet \cite{bonacich1987power,brin1998anatomy,kleinberg1999authoritative,page1999pagerank}.
More recently, the class of matrix function-based centrality measures has attracted significant attention \cite{katz1953new,estrada2005subgraph,estrada2010network,benzi2013total,benzi2020matrix}.
One distinguishing feature of these measures is their flexibility to interpolate between the well-established local degree and global eigenvector centrality measures \cite{benzi2015limiting}.
Recent contributions have generalized eigenvector and matrix function-based centrality measures to multiplex networks, a special class of multilayer networks in which edges across layers are only present between nodes representing the same entity in different layers, cf.\ e.g., \cite{taylor2017eigenvector,bergermann2021orientations,bergermann2022fast}.

Community detection methods, instead, aim at identifying groups of densely connected entities in complex networks that are sparsely connected to other communities \cite{girvan2002community,newman2006modularity,von2007tutorial,fortunato2010community}.
One successful concept to obtain such partitions is the maximization of modularity, which compares a given connectivity structure with an idealized community structure described by some null model \cite{girvan2002community,newman2006modularity,blondel2008fast}.
The first null model for multilayer networks has been proposed a number of years ago \cite{mucha2010community}.

The qualitative analysis of this paper is based on Foucault's discourse term of knowledge and power \cite{foucault1970archaeology} as well as Berger and Luckmann's concept of social constructivism \cite{berger1967social}.
The sociology of knowledge approach to discourse (SKAD) by Keller combines these two theoretical approaches into the methodological framework employed in this paper \cite{keller2011sociology}.
We analyze the content structure in terms of the phenomenal structure of the tweets as well as the subject relations in a chronological analysis \cite{keller2011sociology}.
Recent studies using discourse analysis approaches have examined political, corporate, and activist communication on other events such as COVID-19 \cite{wicke2021covid}, a Spanish journalist scandal \cite{reguero2023journalism}, or the 2016 US election \cite{yaqub2017analysis}.
Further discourse analyses focus on different communication strategies of actor groups such as NGOs \cite{waters2011tweet}, activists \cite{shea2022david}, or brands \cite{wu2018emotional}.

With the theoretical background discussed above, we understand the quantitative analysis as the reconstruction of subject positions, their interactions, and community formation, serving as an example how qualitative and quantitative methodology, e.g., mixed methods complement each other.

In this work, we combine quantitative network-theoretic and qualitative discourse-theoretic methodology for the analysis of the global Twitter discourse on the Rana Plaza factory collapse.
We separately analyze three week time windows around the collapse (anniversary) date over ten years, which allows for the detection of the temporal evolution of the discourse.

The remainder of this paper is organized as follows.
\Cref{sec:data} describes the data collection procedure.
In \Cref{sec:methods}, we introduce the quantitative network-theoretic and qualitative discourse-theoretic methods employed in this work.
\Cref{sec:results} describes the results of the analysis before we discuss our findings in \Cref{sec:discussion} and summarize in \Cref{sec:conclusion}.

\section{Data}\label{sec:data}

The analysis of this study is based on tweet data of the Twitter discourse on the collapse of the Rana Plaza textile factory in Dhaka, Bangladesh on April $24$th $2013$.
The data was obtained from the Twitter API with a tweet text filter for the term ``Rana Plaza''.
The tweets come in \texttt{json} format containing tweet text, tweet time, tweet ID, user ID, and information on interactions such as retweets, replies, and mentions as mandatory platform-generated fields.
Further fields such as user names, locations, descriptions, etc.\ are mostly optional and provided by the individual users.

We collected data for the ten years from $2013$ in which the event took place until $2022$.
Each year covers the same time window from April $17$th to May $10$th around the (anniversary) date April $24$th.
We refrained from manual data processing, which leaves the possibility of encountering (in practice very rare) off-topic tweet content.

Our search procedure found a total number of $101\,103$ tweets within the time windows detailed above.
The tweet numbers are inhomogeneously distributed across the ten years with $15\,522$ tweets in $2013$ and $22\,322$ tweets in $2014$ marking the two highest tweet numbers per year.
The lowest number of $4\,888$ tweets per year was encountered in year $2019$.
\Cref{fig:tweet_count} illustrates the total yearly tweet volumes as well as yearly tweet numbers for certain countries relevant to this study either due to their high tweet volumes or their geographical proximity to the Rana Plaza factory.
In addition, \Cref{fig:daily_tweet_volume} provides daily tweet volumes for all years over time intervals of two hours.

The analysis of data from discourse participants from all over the world provides an overview on the international community structure, central users, and central, thus potentially contentually relevant tweets.

\section{Methods}\label{sec:methods}

This section introduces existing methodology employed to obtain the results of this study.
The first illustrations in \Cref{sec:results_tweets} for exploratory data analysis are obtained by straightforward aggregation of tweet numbers over varying time intervals.
The word frequency analysis displayed in \Cref{tab:word_frequency_bars} and discussed in \Cref{sec:discussion} is based on a word combination search with two to four words as well as a single word frequency search in MaxQDA, excluding words such as ``a'' or ``the''.
\Cref{sec:methods_multiplex} describes the quantitative network-theoretic and \Cref{sec:methods_content} the qualitative discourse-theoretic methods employed in this work before \Cref{sec:methods_interplay} describes how their interplay leads to the results presented in \Cref{sec:results}.

\subsection{Quantitative multiplex network analysis}\label{sec:methods_multiplex}

For the quantitative analysis of the network structure of Twitter user interactions we rely on a multiplex network model \cite{kivela2014multilayer,boccaletti2014structure}.
We follow the modeling approach presented by \cite{omodei2015characterizing} and identify layers with different types of user interactions: retweet, reply, and mention.
We construct a separate multiplex network for each year, which allows a comparison of different structural network properties across the years.

The basis of our network model is the identification of users with nodes and user interactions with edges between pairs of nodes.
We use weighted directed edges as each interaction (retweet, reply, and mention) involves two users clearly distinguishable as initiator and recipient.
Edge weights correspond to the numbers of each type of interaction between all pairs of users per year.

To construct the network, we define the node (or vertex) set $\mathcal{V}$ that consists of all $n$ users participating in the Twitter discourse of the respective year.
Furthermore, we define three intra-layer edge sets $\mathcal{E}^{(1)}, \mathcal{E}^{(2)},$ and $\mathcal{E}^{(3)}$ denoting the retweet, reply, and mention interactions, respectively.
Each of the $L=3$ layers is represented by a generally nonsymmetric adjacency matrix $\bm{A}^{(l)}\in\mathbb{R}^{n \times n}$ with entries
\begin{equation*}
\bm{A}^{(l)}_{ij} = 
\begin{cases}
w^{(l)}_{ij} & \text{if interaction $l$ took place between users $i$ and $j$,}\\
0 & \text{otherwise,}
\end{cases}
\end{equation*}
for $l=1, 2, 3,$ where $w^{(l)}_{ij}$ denotes the number of distinct interactions $l$ initiated by user $i$ and received by user $j$, e.g., if user $i$ retweets seven tweets by user $j$ we have $\bm{A}^{(1)}_{ij} = 7$.
The final ingredient that completes our multiplex network $\mathcal{G} = (\mathcal{V}, \mathcal{E}^{(1)}, \mathcal{E}^{(2)}, \mathcal{E}^{(3)}, \tilde{\mathcal{E}})$ is the inter-layer edge set $\tilde{\mathcal{E}}$.
It represents interactions of nodes from different layers and in this work we use diagonal all-to-all coupling, i.e., each node is connected to itself in both other layers with weight $1$.

\begin{figure}
	\begin{center}
\subfloat[$\bm{A}^{(1)}$, retweet]{
	\includegraphics[width=.33\textwidth]{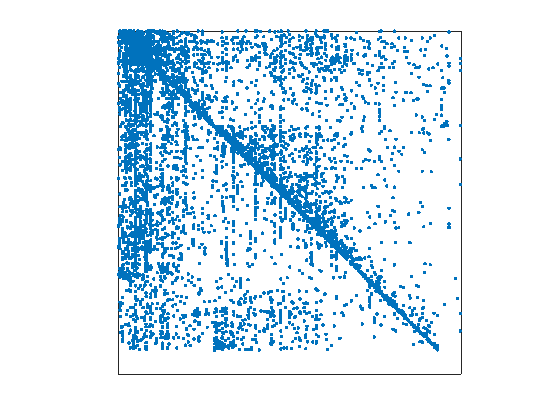}
}
\subfloat[$\bm{A}^{(2)}$, reply]{
\includegraphics[width=.33\textwidth]{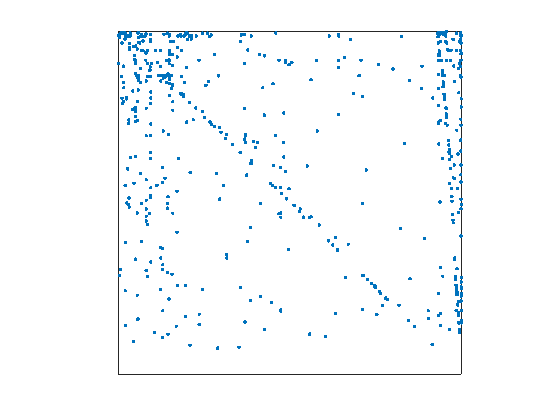}
}
\subfloat[$\bm{A}^{(3)}$, mention]{
\includegraphics[width=.33\textwidth]{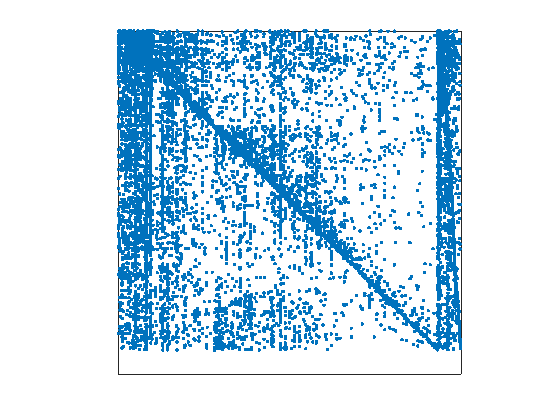}
}
\end{center}
\caption{Sparsity structure of the intra-layer adjacency matrices of year $2014$.}\label{fig:example_adjacencies}
\end{figure}

For the quantitative analysis, we use the linear algebraic representation of the multiplex network $\mathcal{G}$ in terms of the supra-adjacency matrix $\bm{A}\in\mathbb{R}^{nL \times nL}$ defined as

\begin{equation}\label{eq:supra_adjacency}
\bm{A} = \bm{A}_{\mathrm{intra}} + \omega \bm{A}_{\mathrm{inter}} = 
\begin{bmatrix}
\bm{A}^{(1)} & \bm{0} & \bm{0}\\
\bm{0} & \bm{A}^{(2)} & \bm{0}\\
\bm{0} & \bm{0} & \bm{A}^{(3)}
\end{bmatrix}
+ 
\omega
\begin{bmatrix}
\bm{0} & \bm{I} & \bm{I}\\
\bm{I} & \bm{0} & \bm{I}\\
\bm{I} & \bm{I} & \bm{0}
\end{bmatrix},
\end{equation}
where $\omega\in\mathbb{R}$ denotes a scalar layer-coupling parameter, $\bm{0}\in\mathbb{R}^{n \times n}$ the matrix of all zeros, and $\bm{I}\in\mathbb{R}^{n \times n}$ the identity matrix.
Throughout our analysis, we choose the layer-coupling parameter $\omega=1$.
As an illustrative example, \Cref{fig:example_adjacencies} shows the intra-layer sparsity structure of the three layers for the year $2014$.
The blank lower blocks in each layer indicate a set of users that receives user interaction without initiating it.
The adjacency matrices for all ten years are publicly available in Matlab and python formats \cite{bergermannTwitter}.

\subsubsection{Matrix function-based centrality measures}\label{sec:methods_multiplex_centralities}

The identification and ranking of influential individuals in networked complex systems is an important area of research in network science.
Prominent examples of well-established centrality measures have been developed over past decades \cite{freeman1977set,freeman1978centrality,bonacich1987power,kleinberg1999authoritative} with PageRank at the heart of Google's search engine being an example of a centrality measure impacting everyday life \cite{brin1998anatomy,page1999pagerank}.

A more recently studied class of centrality measures is the class of matrix function-based centrality measures \cite{katz1953new,estrada2005subgraph,estrada2010network,benzi2013total,benzi2020matrix}.
These can be interpreted as walk-based measures taking all possible connections between all pairs of nodes of the network into account.
Matrix function-based centrality measures possess the flexibility to interpolate between local degree and global eigenvector centrality \cite{benzi2015limiting} by means of a parameter called $\alpha$ or $\beta$ that controls how strongly longer walks are weighted.

In this work, we employ a recently introduced generalization of matrix function-based centrality measures to multiplex networks \cite{bergermann2021orientations,bergermann2022fast} for the identification of key actors in the Twitter discourse under investigation.
We restrict ourselves to the multiplex version of receiver total communicability \cite{benzi2013total} defined as
\begin{equation}\label{eq:def_tc}
\mathrm{TC}(i,\beta) = \bm{e}_i^T \exp(\beta\bm{A}^T) \bm{1} = \bm{e}_i^T \left( \sum_{p=0}^\infty \frac{\beta^p}{p!} (\bm{A}^T)^p \right) \bm{1},
\end{equation}
and receiver Katz centrality \cite{katz1953new} defined as
\begin{equation}\label{eq:def_kc}
\mathrm{KC}(i,\alpha) = \bm{e}_i^T (\bm{I} - \alpha\bm{A}^T)^{-1} \bm{1} = \bm{e}_i^T \left( \sum_{p=0}^\infty \alpha^p (\bm{A}^T)^p \right) \bm{1},
\end{equation}
where $\beta > 0$ and $0 < \alpha < 1/\lambda_{\mathrm{max}}$ with $\lambda_{\mathrm{max}}$ the largest eigenvalue of $\bm{A}$ denote the trade-off parameters mentioned above.
Furthermore, $\bm{e}_i\in\mathbb{R}^{nL}$ denotes the $i$th canonical basis vector, $\bm{1}\in\mathbb{R}^{nL}$ the vector of all ones, and $\exp(\beta\bm{A}^T)$ and $(\bm{I} - \alpha\bm{A}^T)^{-1}$ defined by the above matrix power series denote the matrix exponential and resolvent function, respectively \cite{higham2008functions}.

In terms of the Twitter interaction networks at hand, these measures can be interpreted as simulating the effectivity of the distribution of information along existing channels in the network.
By the choice of total communicability and Katz centrality, \eqref{eq:def_tc} and \eqref{eq:def_kc} measure the impact that all information distributed by node $i$ has on the entirety of the network.
The choice of receiver centralities leads to high rankings of users being retweeted, being replied to, and being mentioned and causes the transpose of $\bm{A}$ in the above definitions.

For the computation of the vectors $\exp(\beta\bm{A}^T) \bm{1}$ and $(\bm{I} - \alpha\bm{A}^T)^{-1} \bm{1}$, which contain the measures \eqref{eq:def_tc} and \eqref{eq:def_kc} for all nodes $i=1, \dots , nL$, we employ highly efficient methods from numerical linear algebra, which are described in detail in \cite{bergermann2022fast} and implemented in Matlab \cite{bergermannMMFC} and python \cite{bergermannUrban}.
The centrality results of this paper can be reproduced using publicly available codes \cite{bergermannTwitter}.

The quantities ultimately used in our analysis and displayed in Figure~\ref{fig:centralities_communities} are one TC and one KC centrality value for each Twitter user per year.
These are easily obtained from \eqref{eq:def_tc} and \eqref{eq:def_kc} by marginal node centralities, i.e., the addition of each node's centralities from the three layers \cite{taylor2017eigenvector,bergermann2022fast,bergermann2021orientations}.
Throughout our computations, we use the parameters $\alpha=0.5/\lambda_{\mathrm{max}}$ and $\beta=2/\lambda_{\mathrm{max}}$.

\subsubsection{The Louvain community detection method}\label{sec:methods_multiplex_communities}

A second structural network feature of tremendous interest to network science is community detection \cite{girvan2002community,newman2006modularity,von2007tutorial,fortunato2010community}.
Similar to the small-world effect, scale-freeness, or transitivity, the presence of a community structure is viewed as a universal property of networks originating from various disciplines including social networks \cite{watts1998collective,newman2003structure}.

A community in a network is generally understood as a subset of nodes that is strongly connected internally but only sparsely connected to the remainder of the network.
One widely acknowledged concept to quantify this property is network modularity \cite{girvan2002community,newman2006modularity}.
It compares a given distribution of edges in an observed network with an expected distribution according to some null model and defines a scalar measure $Q$ of the quality of a given partition of the node set.
Optimal partitions can hence be obtained by computational maximization of the modularity score $Q$.

One successful heuristic computational method for modularity maximization is the ``Louvain method'' \cite{blondel2008fast}.
It initially assigns every node its own community and proceeds by iteratively merging pairs of communities that lead to a maximal increase in the modularity $Q$.
The method terminates if no further improvement of the partition is possible.

A generalization of the Louvain method to multilayer networks was proposed by \cite{mucha2010community}.
It defines a multilayer null model that in our notation leads to the multilayer modularity function
\begin{equation}
Q = \frac{1}{2\mu} \sum_{ijlr} \left[ \left( \bm{A}_{ij}^{(l)} - \gamma_l \frac{k_{il}k_{jl}}{2m_l} \right) \delta_{lr} + \omega\delta_{ij} \right] \delta(g_{il},g_{jr})
\end{equation}
for a given node partition $g$, where $g_{il}$ denotes the community affiliation of node $i$ in layer $l$, $\gamma_l$ denotes a resolution parameter, and $\delta_{ij}, \delta_{lr},$ and $\delta(g_{il},g_{jr})$ denote Kronecker deltas.
Furthermore, the quantities $2\mu = \sum_{ijl} \bm{A}_{ij}^{(l)} + 6n\omega, k_{il}=\sum_j \bm{A}_{ij}^{(l)}, k_{jl}=\sum_i \bm{A}_{ij}^{(l)}$, and $m_l = \sum_{ij} \bm{A}_{ij}^{(l)}$ denote different strengths of the multiplex network.

We used the Matlab implementation \cite{JeubGenLouvain} for obtaining the community structures reported in \Cref{sec:results_multiplex} and illustrated in Figure~\ref{fig:centralities_communities}.
The results can be reproduced with our code release \cite{bergermannTwitter} together with \cite{JeubGenLouvain}.
The output of the method consists of three community affiliations for each user corresponding to the three layers.
Due to the inter-layer connectivity, the three affiliations mostly coincide.
In case of deviations, a majority vote is performed over the layers, i.e., the most frequently encountered community determines the user's community affiliation.
Throughout our computations, we use the parameter $\gamma_l=1$.

\subsection{Qualitative analysis by the sociology of knowledge approach to discourse}\label{sec:methods_content}

The selection of tweets considered in the qualitative content analysis builds on the centrality measures introduced in \Cref{sec:methods_multiplex_centralities} as we consider all tweets by the union of the ten most central users identified by both centrality measures per year.

For the qualitative analysis of these tweets, we use the sociology of knowledge approach to discourse (SKAD), which is a methodological framework focusing on questions on public discourse arenas and knowledge power structures \cite{keller2011sociology}.
Building on the power and knowledge correlation described by Foucault \cite{foucault1970archaeology} and the power structure in knowledge societies \cite{berger1967social}, the sociology of knowledge approach offers the opportunity to analyze knowledge politics.
Foucault understands discourses as social practices that imply controversies and discoursive fights around ``problemetizations'' \cite{foucault1970archaeology,foucault1997polemics}.
In order to analyze discoursive formations such as the formation of concepts or speaker positions, concrete data, such as in our case tweets, have to be analyzed bottom up \cite{foucault2005order,keller2011sociology}.

In the qualitative analysis of the Twitter discourse, we focus on the phenomenal structure as a content-structuring framework.
The anaysis of content structure allows us to determine the change of content in the tweets over the years and which dimensions of the event are communicated. 
Following the SKAD, we understand phenomenal structures as the construction of an issue and its dimensions such as the definition, causal relations, corresponding responsibilities, normative judgments, and possible courses of action \cite{keller2011sociology}.
As described by Berger and Luckmann \cite{berger1967social}, these dimensions are constructed from individual subject positions that do ``not describe any essential qualities of a discourse topic'' \cite{keller2011sociology} but rather a communicative construction of reality \cite{knoblauch2019communicative}.

The basis of the bottom-up analysis in the SKAD is a coding process, which builds on the methodology of the grounded theory devised by \cite{glaser1967discovery}.
In the coding phase, the raw data is firstly broken down into codes to prepare it for the analysis.
Each code is assigned to a specific passage in the data and multiple codes for a given passage are possible.
In this work, each tweet is coded with at least one code, e.g., ``\textit{8 years ago, the Rana Plaza collapse showed that voluntary agreements don't keep workers safe. Then why haven't brands comit[ted]...}'' is coded with both \texttt{Memorial} and \texttt{Security standards/Worker's rights}.
In the coding process, we used in-vivo codes, i.e., codes taken directly from the data as well as contrasting and association techniques to identify the initial codes.
For the coding process and the schematic design of \Cref{fig:code_network}, we used MaxQDA 2020.

Secondly, the codes are revised for the dimensions introduced above as well as relationships between them, i.e., codes can be merged or diversified.
To carve out the content structure, the interpretative analysis process is oriented towards the problem dimensions of the phenomenal structure, \cite{keller2011sociology} describes.

The resulting dimensions of our study alongside corresponding codes (where applicable) are summarized in \Cref{tab:code_overview}.
Each code and dimension is accompanied by a short description as well as an example tweet.
The links and relations of the dimensions identified in this work are depicted in \Cref{fig:code_network}.
We particularly focus on the communicators' subject positions in the discourse, which can be sub-divided into the user's nationalities, which are summarized in \Cref{fig:tweet_count}, and their actor groups.
We identified the five actor groups ``Activist'', ``Politician'', ``News agency'', ``Anonymous user'', and ``Brand''.
Due to the frequent occurrence of the actor group ``Activist'', this group was sub-divided into ``NGO'' (Non-Governmental Organization), ``Writer'', and ``Celebrity''.

\subsection{Interplay of quantitative and qualitative methods}\label{sec:methods_interplay}

The starting point of our analysis is the abstraction of Twitter interactions into multiplex networks as described in \Cref{sec:methods_multiplex} followed by the centrality analysis presented in \Cref{sec:methods_multiplex_centralities}.
All tweets of the union of the ten most central users of both centrality measures per year present the basis for the qualitative analysis by the SKAD discussed in \Cref{sec:methods_content}.
During the coding process, each central user is manually reviewed and assigned to an actor group.
These actor groups in turn serve as means to visualize the community detection results obtained by the multilayer Louvain method described in \Cref{sec:methods_multiplex_communities}.
As most communities are found to gather around one distinctly most central user we identify each community with the actor group affiliation of that most central user.
This allows to highlight the differing composition and heterogeneous interconnectedness of communities illustrated in Figure~\ref{fig:centralities_communities} through the years.

\section{Results}\label{sec:results}

\subsection{Tweet statistics}\label{sec:results_tweets}

\begin{figure}
	\begin{center}
	\includegraphics[width=0.9\textwidth]{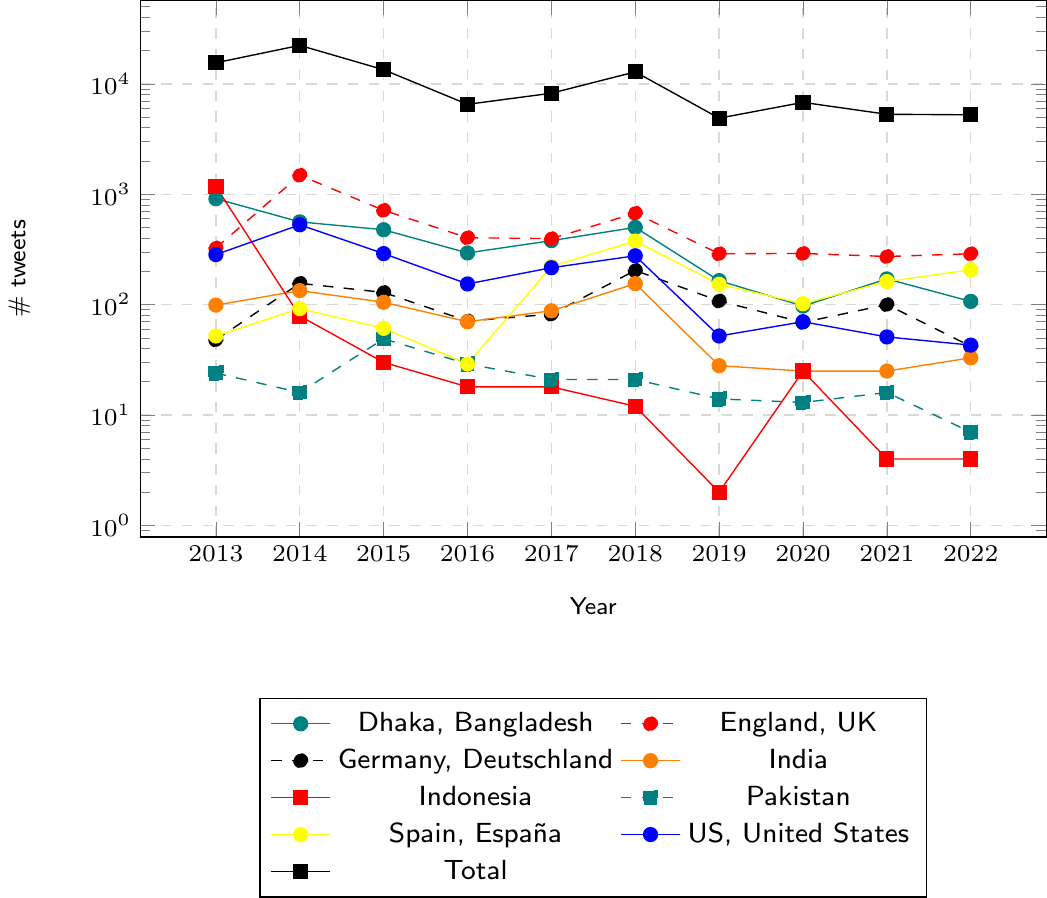}
	\end{center}
	\caption{Yearly tweet volumes for selected countries and in total.
	User-specified locations were filtered by up to two search terms indicated in the legend.
	The total number corresponds to all tweets matching the search term ``Rana Plaza'' in the given time period, cf.~\Cref{sec:data}.}\label{fig:tweet_count}
\end{figure}

We start by reporting some basic numbers on the tweet volume obtained by our search queries described in \Cref{sec:data}.
\Cref{fig:tweet_count} shows that the total number of tweets per year varies between $4\,900$ in year $2019$ and $22\,300$ in year $2014$.
Furthermore, we selected eight countries relevant either due to their high tweet volume or their geographical proximity to the Rana Plaza factory and report their tweet volumes in \Cref{fig:tweet_count}.
We remark that the user location is a user-specified free text field that can not be expected to provide complete information.
While the observed tweet volume in some countries behaves approximately proportionally to the total volume, we observe an increase in tweet volumes from Spain starting from $2017$.
Furthermore, Indonesia shows a remarkably high tweet volume in the year of the collapse, $2013$, followed by a rapid decline and very low tweet volumes in subsequent years.

\begin{figure}
	\begin{center}
	\subfloat[$2013$]{
		\includegraphics[width=.19\textwidth]{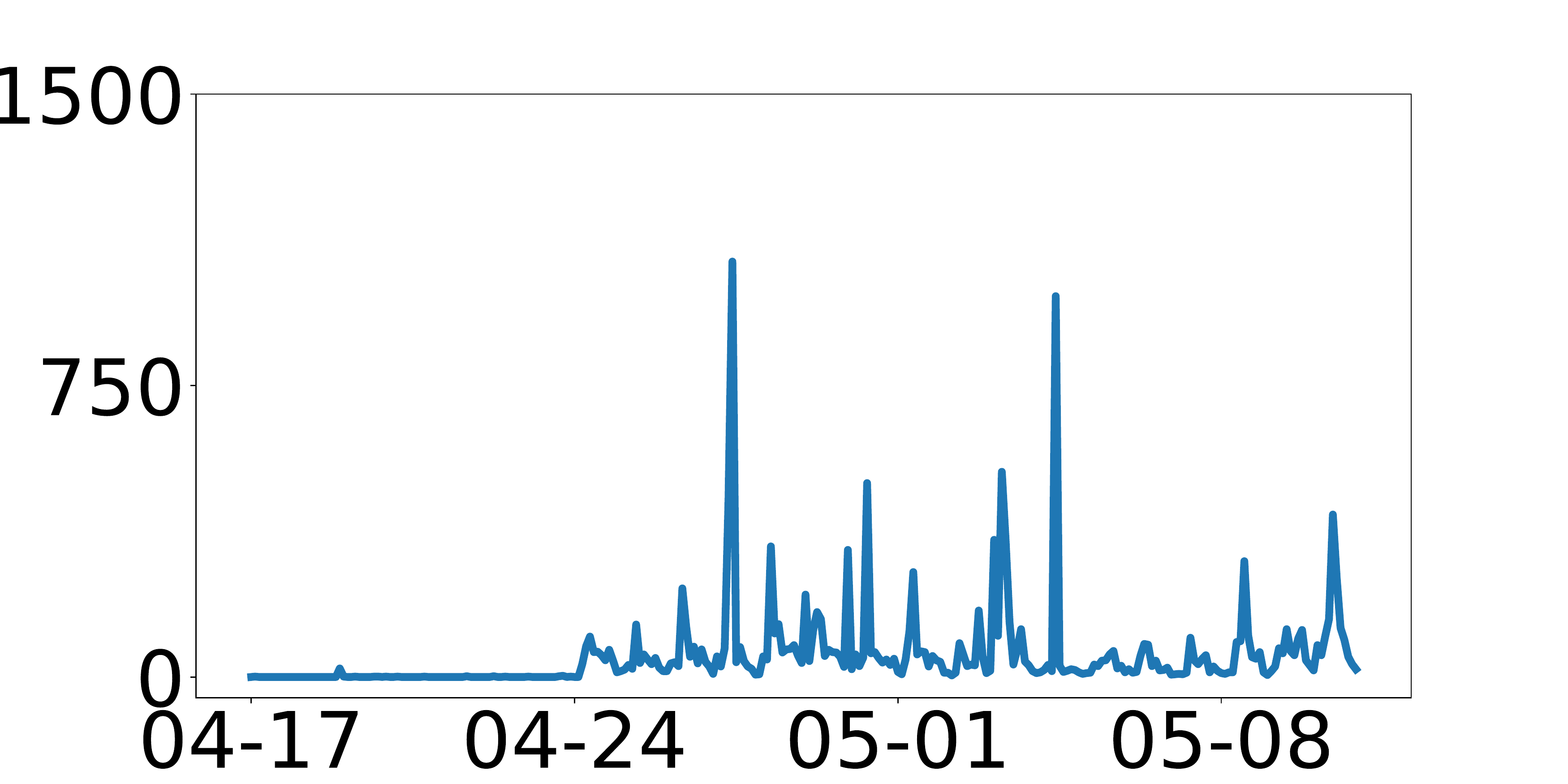}
	}
	\subfloat[$2014$]{
		\includegraphics[width=.19\textwidth]{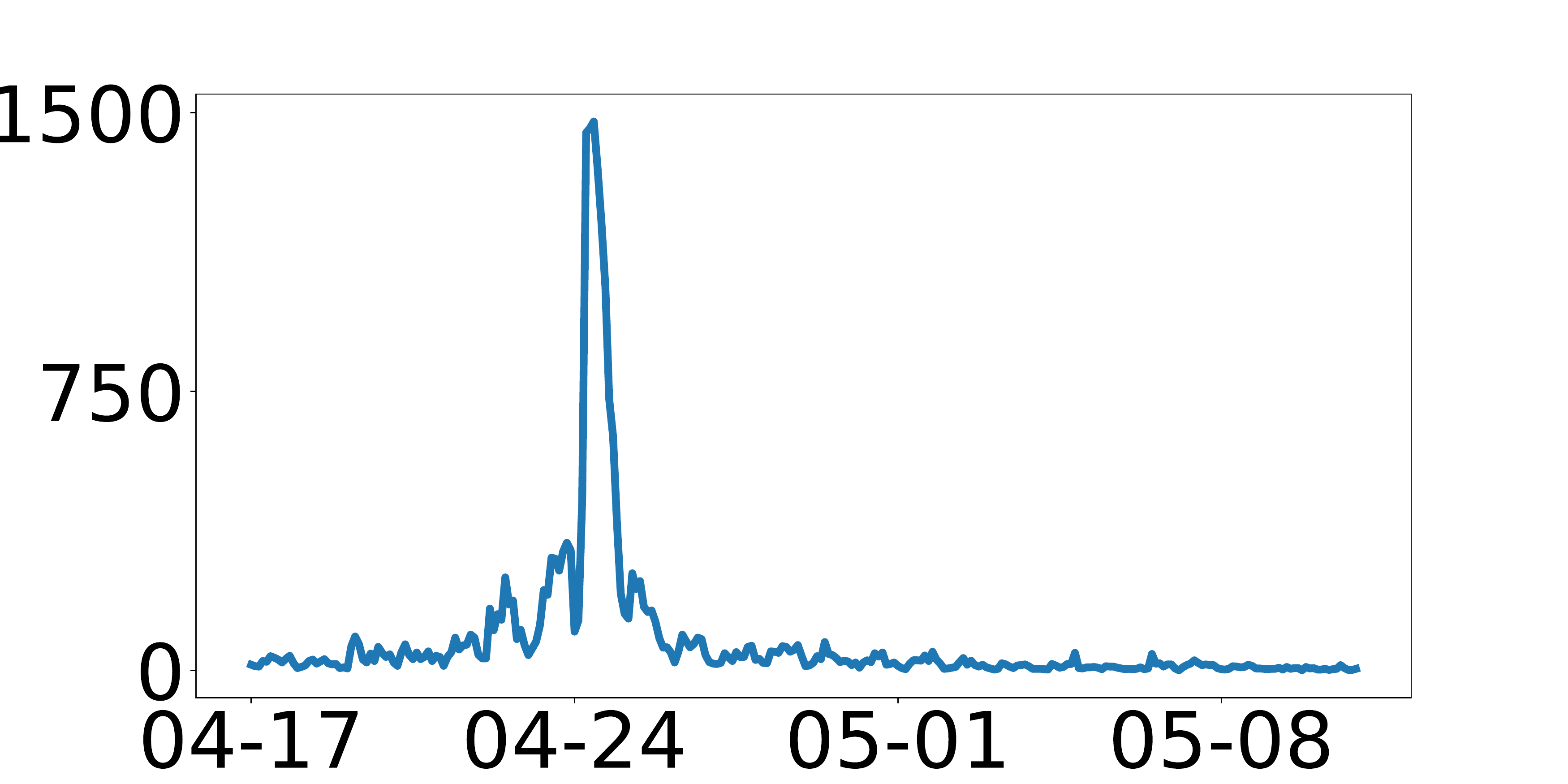}
	}
	\subfloat[$2015$]{
		\includegraphics[width=.19\textwidth]{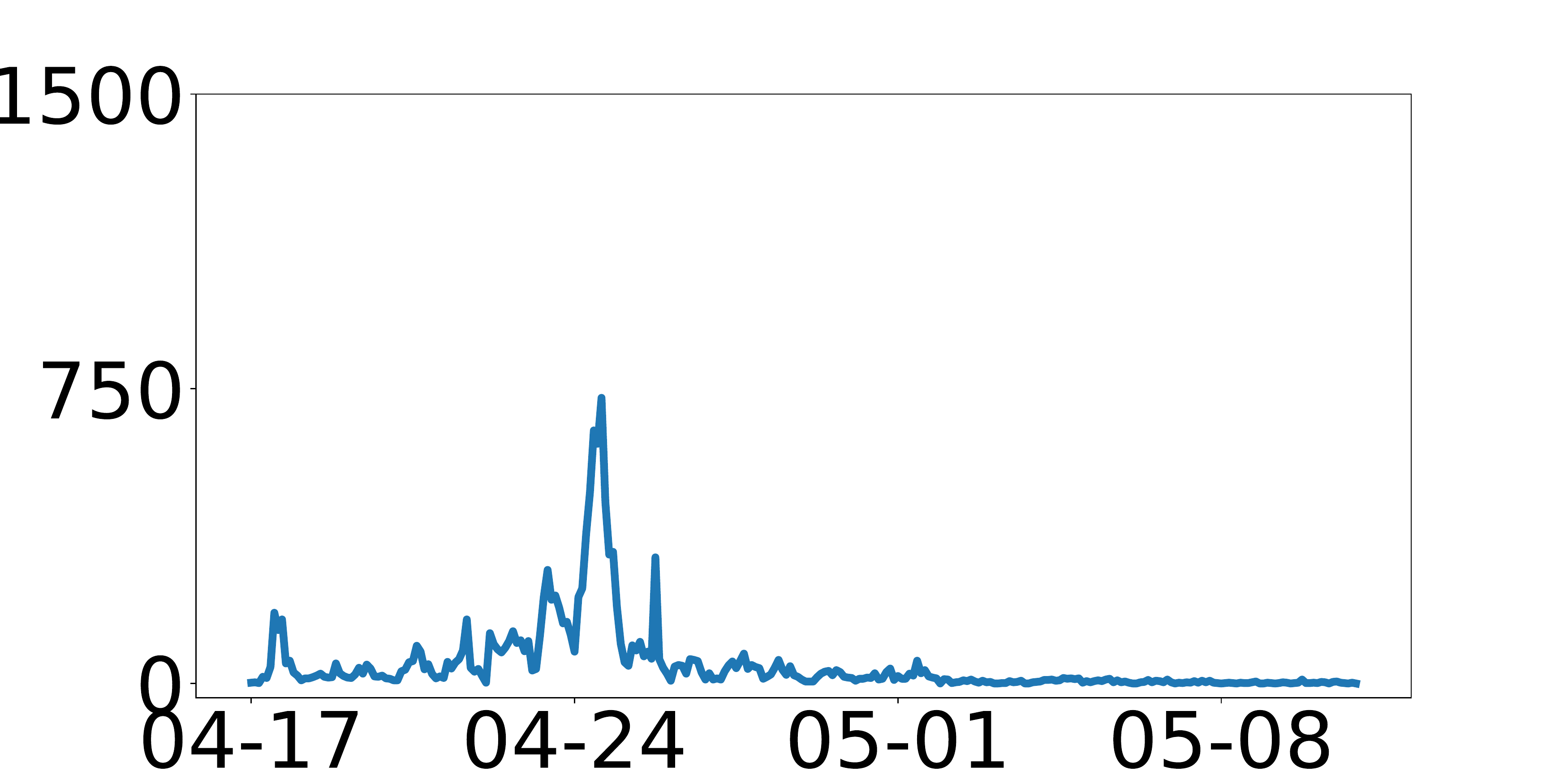}
	}
	\subfloat[$2016$]{
		\includegraphics[width=.19\textwidth]{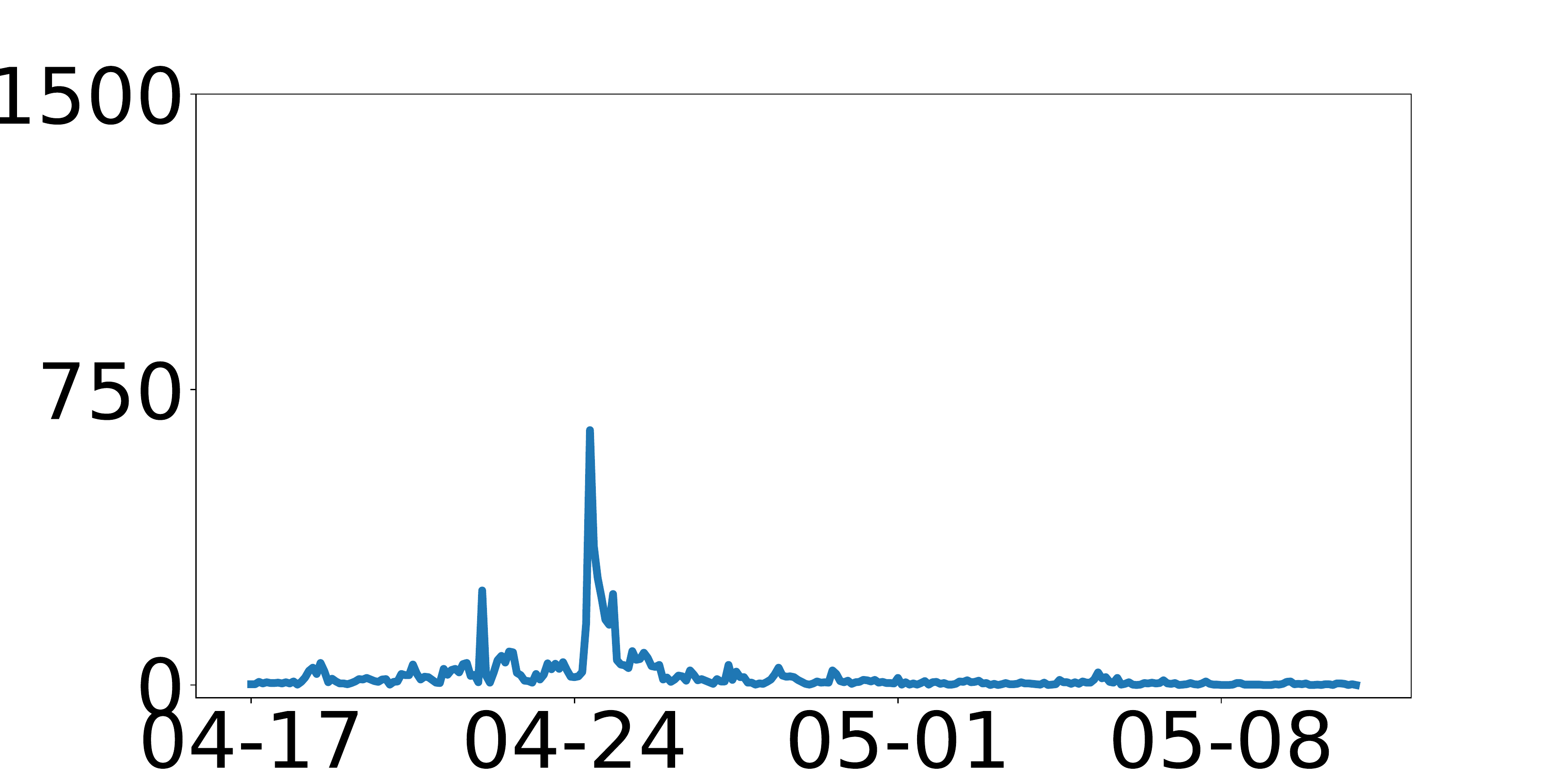}
	}
	\subfloat[$2017$]{
		\includegraphics[width=.19\textwidth]{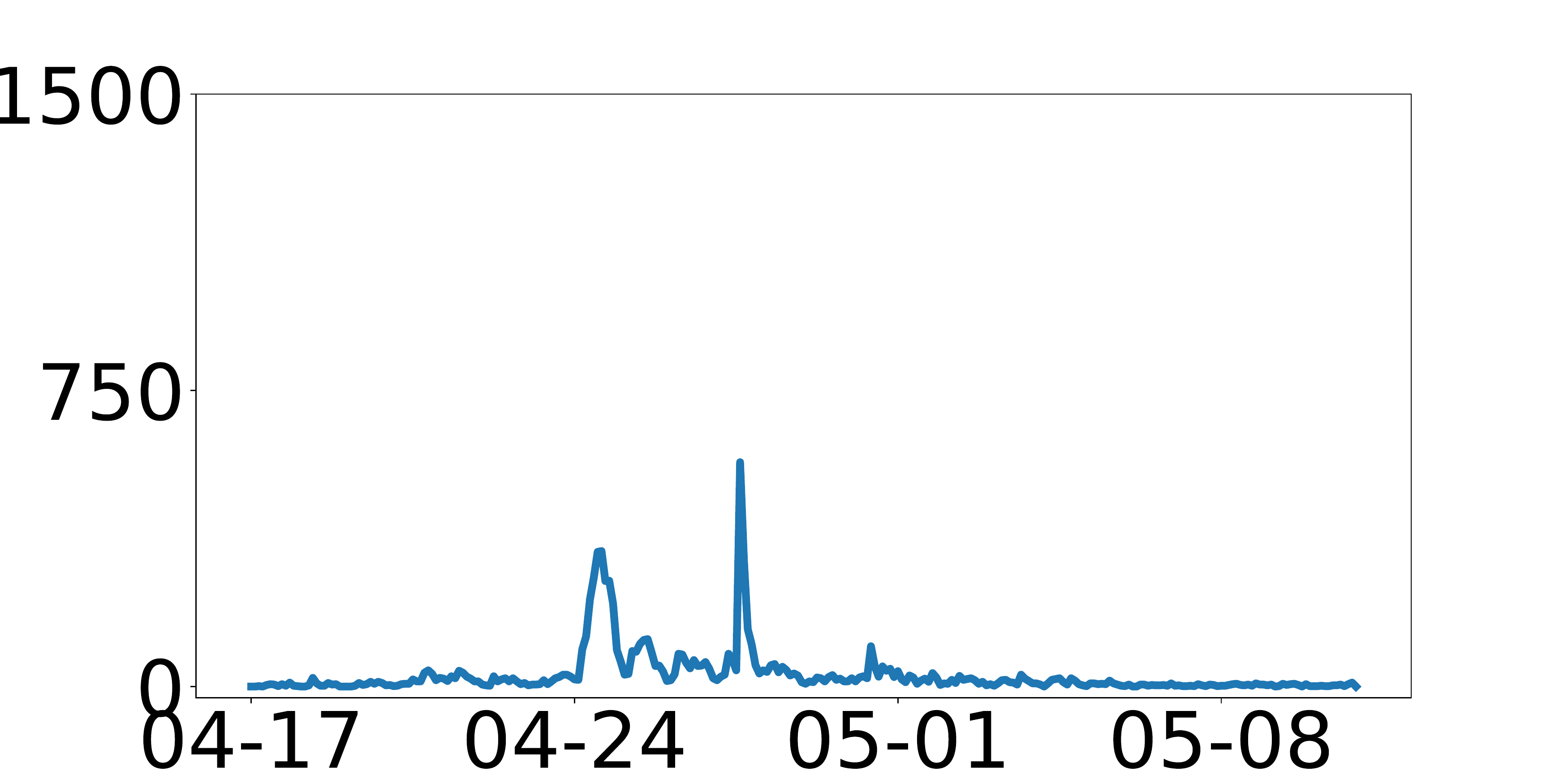}
	}
	
	\subfloat[$2018$]{
		\includegraphics[width=.19\textwidth]{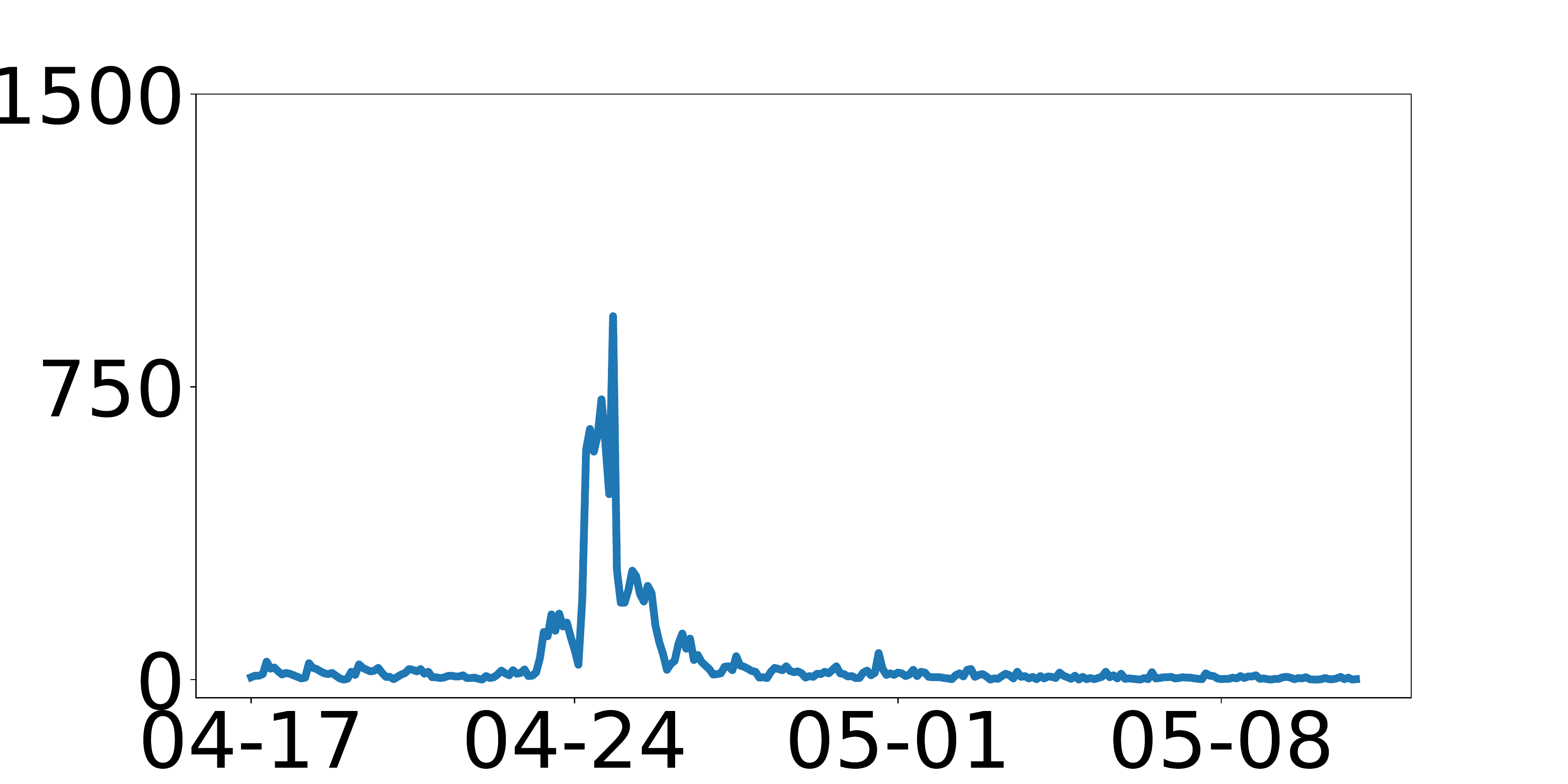}
	}
	\subfloat[$2019$]{
		\includegraphics[width=.19\textwidth]{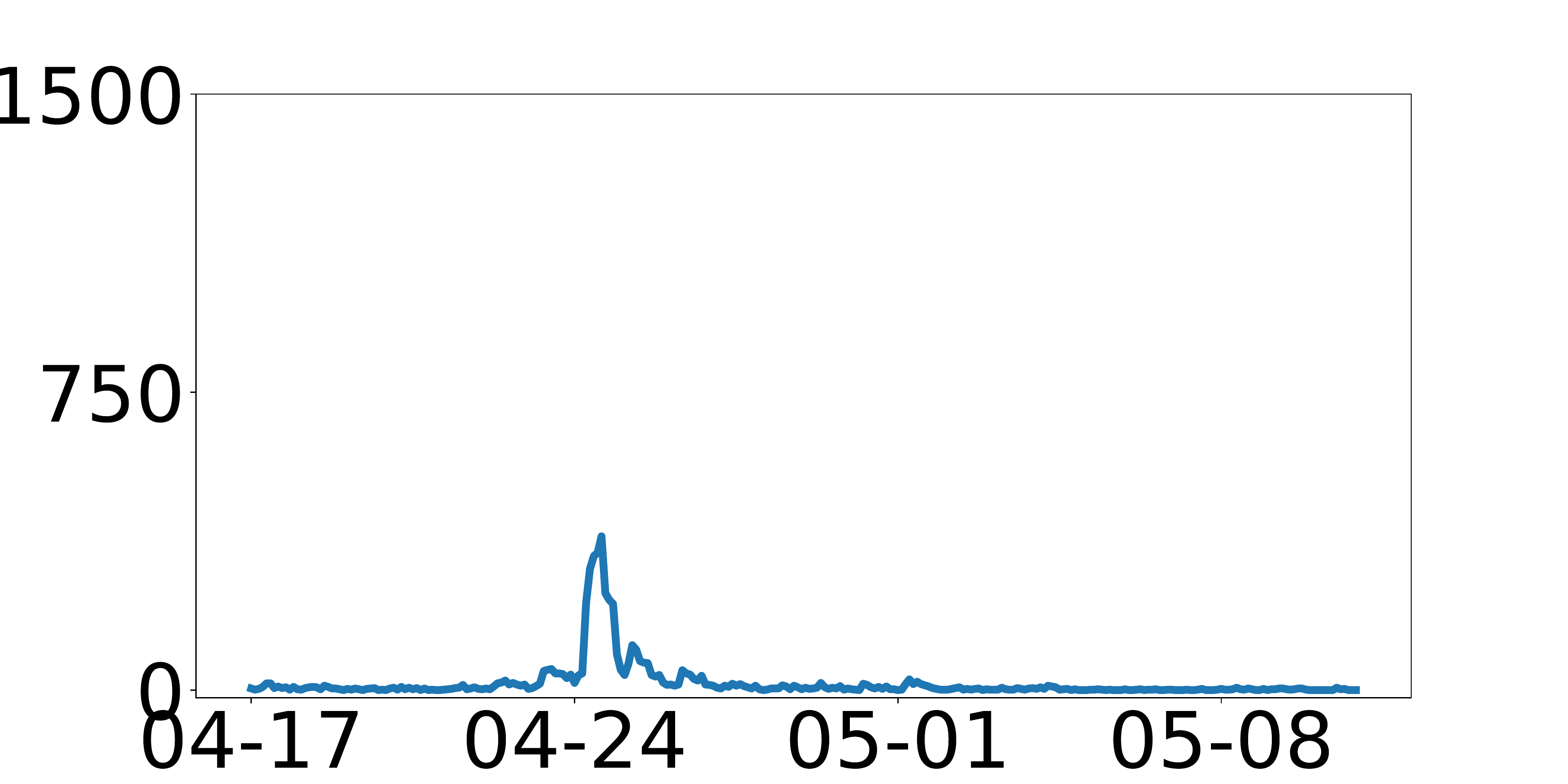}
	}
	\subfloat[$2020$]{
		\includegraphics[width=.19\textwidth]{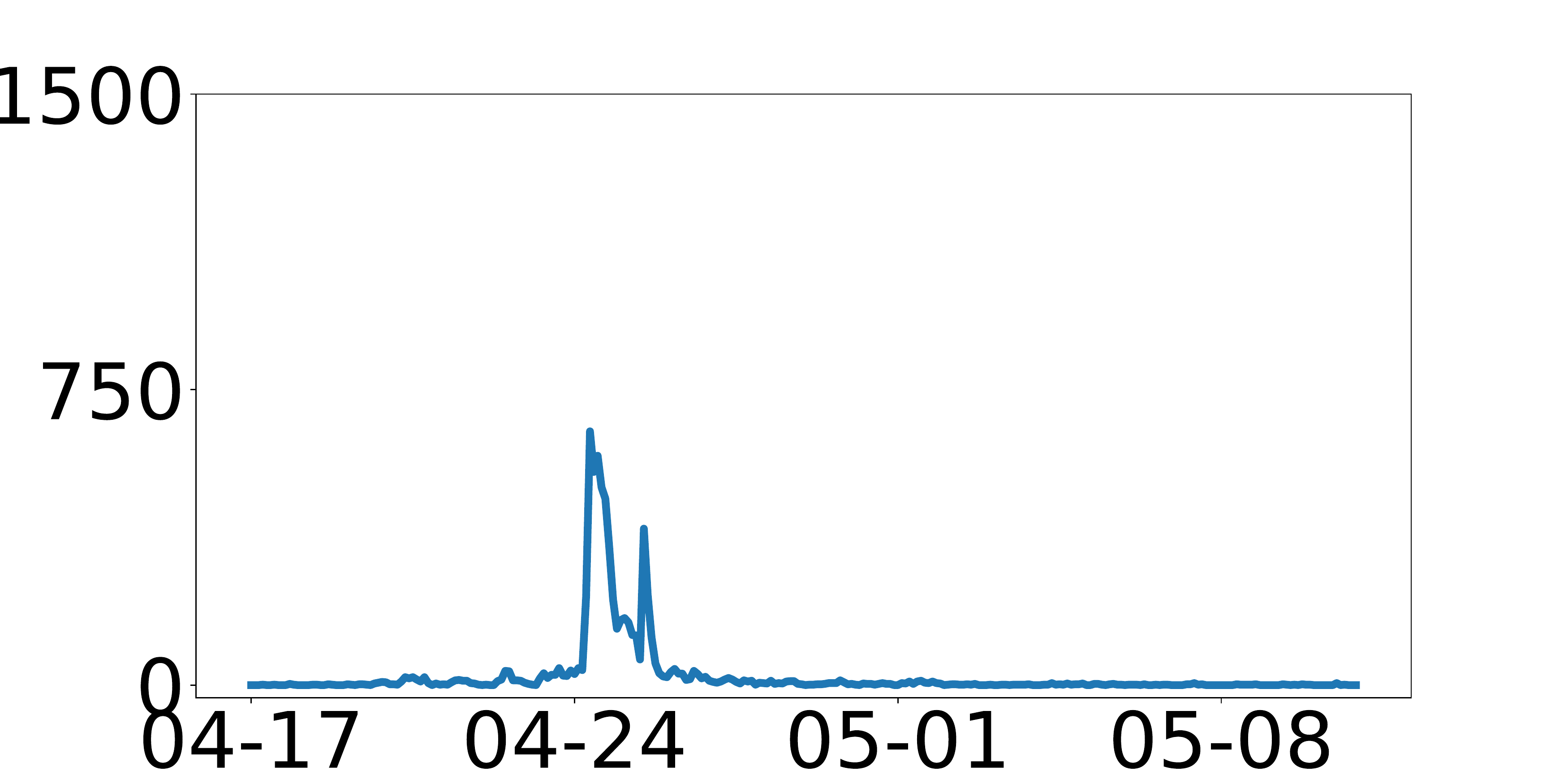}
	}
	\subfloat[$2021$]{
		\includegraphics[width=.19\textwidth]{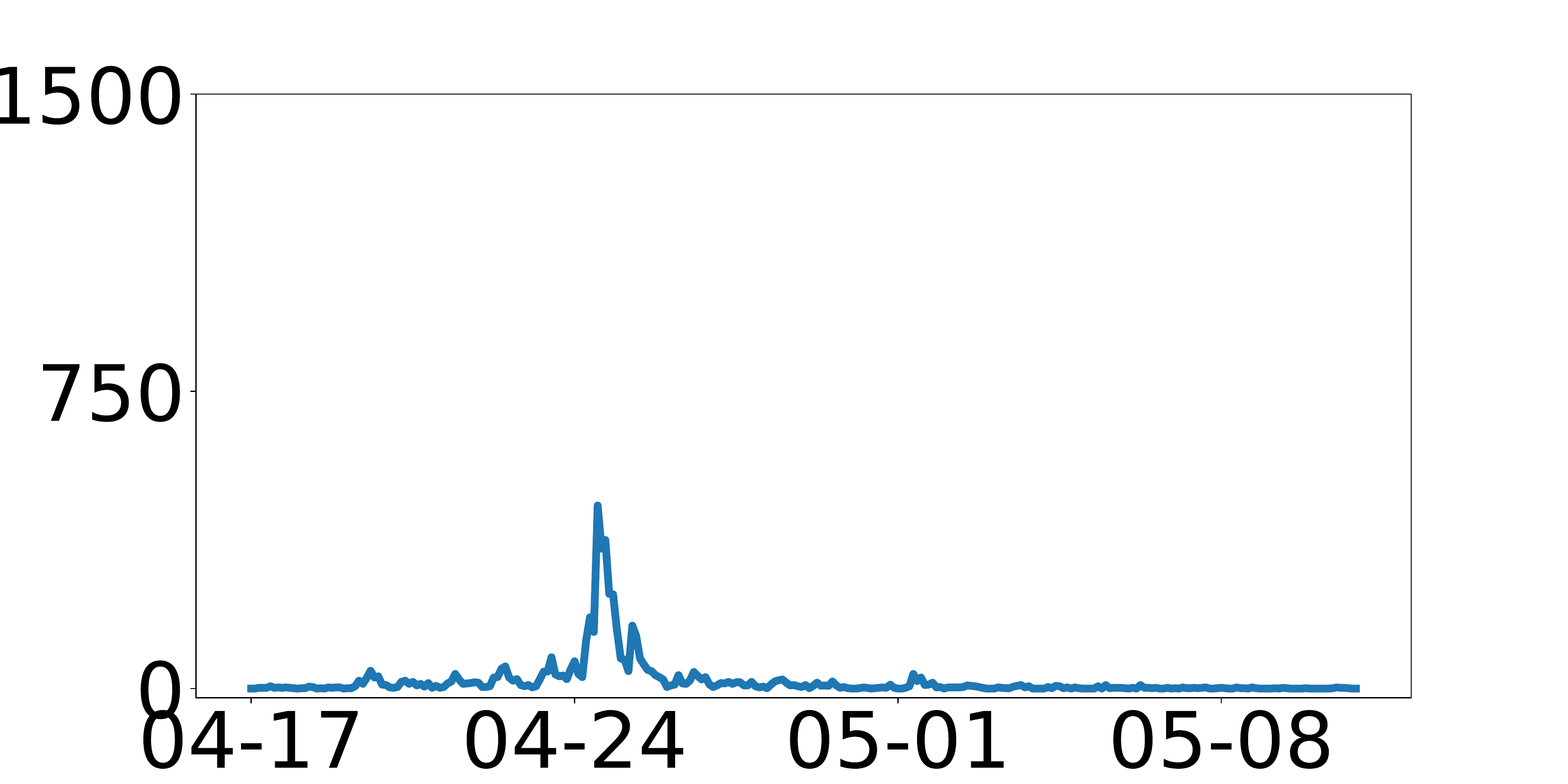}
	}
	\subfloat[$2022$]{
		\includegraphics[width=.19\textwidth]{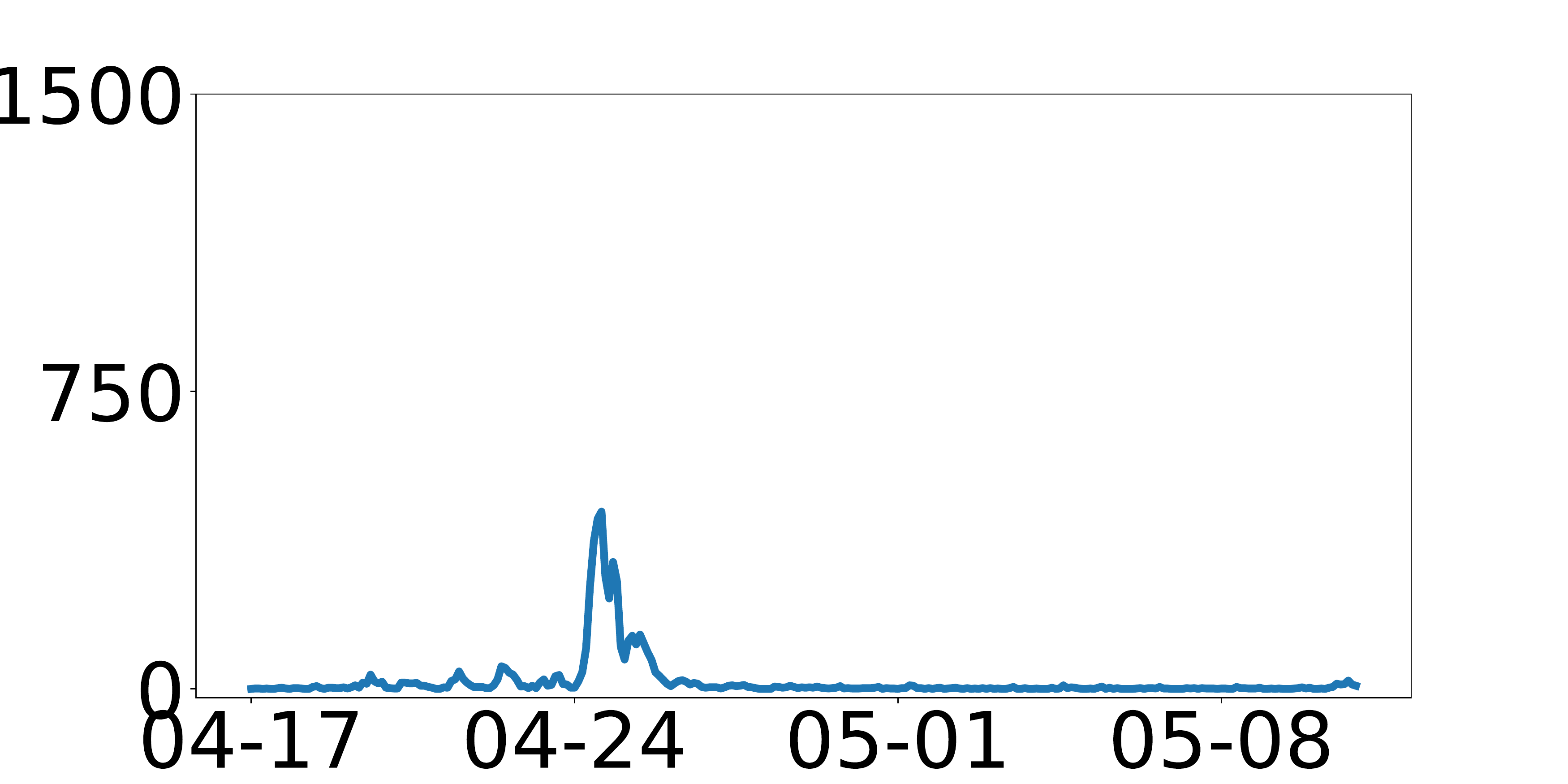}
	}
	\end{center}
	\caption{Tweet volume distribution over two hour time intervals of all considered time periods.
		Abscissa values indicate the time interval from April 17th to May $10$th for the respective year.}\label{fig:daily_tweet_volume}
\end{figure}

\Cref{fig:daily_tweet_volume} additionally shows the distribution of the total tweet volume per year over the respective time interval from April $17$th to May $10$th.
Ordinate values indicate the number of tweets per two
hour time interval.
Naturally, the $2013$ discourse starts with the collapse date April $24$th.
Interestingly, high peak tweet numbers are observed only three days later.
In subsequent years, the discourse temporally closely centers around the anniversary date April 24th.
The only distinct deviation is April 27th 2017, which marks a tweet by ``EmmaWatson'' causing increased user interaction and hence turns the overall downward trend of tweet volumes established between $2014$ and $2016$.

In terms of the three types of interactions retweets, replies, and mentions, we observe across all years that retweets and mentions occur frequently while replies are rare.
The retweet and mention layers in the multiplex framework described in \Cref{sec:methods_multiplex} thus allow for an analysis of the encountered degree distributions:
across all ten years and in line with results from different Twitter discourses on exceptional events \cite{omodei2015characterizing}, we observe a power-law degree distribution with an exponent of approximately $-1.75$.

\subsection{Multiplex network analysis}\label{sec:results_multiplex}

We separately apply multiplex matrix function-based centrality measures and the multilayer Louvain method described in \Cref{sec:methods_multiplex_centralities,sec:methods_multiplex_communities} to the user interaction networks of the ten different years, which allows a structural comparison of the recorded Twitter discourse across the years.

The tables in Figure~\ref{fig:centralities_communities} show the ten most central users of each year according to receiver total communicability (TC) and receiver Katz centrality (KC).
Albeit differences in the numerical values and slight deviations in rankings the results of both measures show a broad overlap.
As is customary in centrality analysis, we observe a localization of both measures \cite{martin2014localization}, i.e., high centrality values for a small portion of the nodes as well as small and similar values for the majority of nodes.
The union of the ten most central nodes of both measures per year forms the basis of the qualitative analysis of this work.

\begin{figure}
	\thisfloatpagestyle{empty}
	\subfloat[Year $2013$]{
		\begin{minipage}{1.1\textwidth}
			\includegraphics[width=.4\textwidth]{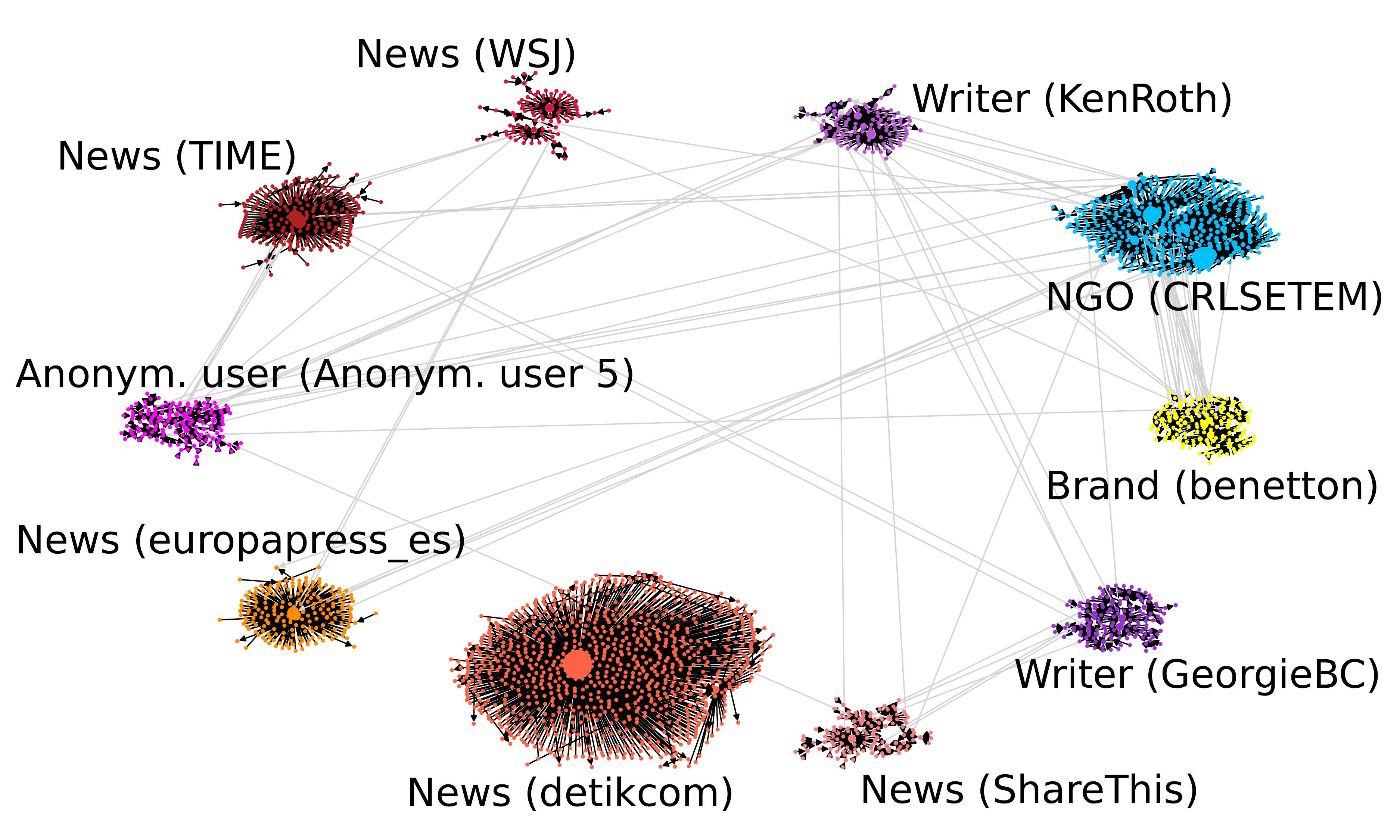}
			\tiny
			\raisebox{44pt}[0pt][0pt]{
				\begin{tabular}{c}
					\hline\hline
					Rank\\\hline
					1\\
					2\\
					3\\
					4\\
					5\\
					6\\
					7\\
					8\\
					9\\
					10\\\hline\hline
				\end{tabular}
				\vspace{5pt}
				\begin{tabular}{lc}
					\hline\hline
					\raisebox{1pt}[0pt][0pt]{\colorbox{white}{}} User & TC\\\hline
					\raisebox{1pt}[0pt][0pt]{\colorbox{tomato}{}} detikcom & 459\\
					\raisebox{1pt}[0pt][0pt]{\colorbox{deepskyblue}{}} CRLSETEM & 271\\
					\raisebox{1pt}[0pt][0pt]{\colorbox{deepskyblue}{}} cleanclothes & 146\\
					\raisebox{1pt}[0pt][0pt]{\colorbox{firebrick}{}} TIME & 102\\
					\raisebox{1pt}[0pt][0pt]{\colorbox{firebrick}{}} TIMEWorld & 89\\
					\raisebox{1pt}[0pt][0pt]{\colorbox{deepskyblue}{}} elcorteingles & 82\\
					\raisebox{1pt}[0pt][0pt]{\colorbox{darkorange}{}} europapress\_es & 62\\
					\raisebox{1pt}[0pt][0pt]{\colorbox{white}{}} BBCBreaking & 49\\
					\raisebox{1pt}[0pt][0pt]{\colorbox{deepskyblue}{}} SETEMPV & 48\\
					\raisebox{1pt}[0pt][0pt]{\colorbox{darkorange}{}} europapress & 47\\\hline\hline
				\end{tabular}
				\vspace{5pt}
				\begin{tabular}{lc}
					\hline\hline
					\raisebox{1pt}[0pt][0pt]{\colorbox{white}{}} User & KC\\\hline
					\raisebox{1pt}[0pt][0pt]{\colorbox{tomato}{}} detikcom & 93.5\\
					\raisebox{1pt}[0pt][0pt]{\colorbox{deepskyblue}{}} CRLSETEM & 38.5\\
					\raisebox{1pt}[0pt][0pt]{\colorbox{deepskyblue}{}} cleanclothes & 27.6\\
					\raisebox{1pt}[0pt][0pt]{\colorbox{firebrick}{}} TIME & 22.6\\
					\raisebox{1pt}[0pt][0pt]{\colorbox{firebrick}{}} TIMEWorld & 18.3\\
					\raisebox{1pt}[0pt][0pt]{\colorbox{darkorange}{}} europapress\_es & 14.6\\
					\raisebox{1pt}[0pt][0pt]{\colorbox{deepskyblue}{}} elcorteingles & 13.2\\
					\raisebox{1pt}[0pt][0pt]{\colorbox{white}{}} BBCBreaking & 12.2\\
					\raisebox{1pt}[0pt][0pt]{\colorbox{darkorange}{}} europapress & 11.8\\
					\raisebox{1pt}[0pt][0pt]{\colorbox{deepskyblue}{}} SETEMPV & 11.2\\\hline\hline
				\end{tabular}
			}
			\normalsize
		\end{minipage}
	}
	
	\subfloat[Year $2014$]{
		\begin{minipage}{1.1\textwidth}
			\includegraphics[width=.4\textwidth]{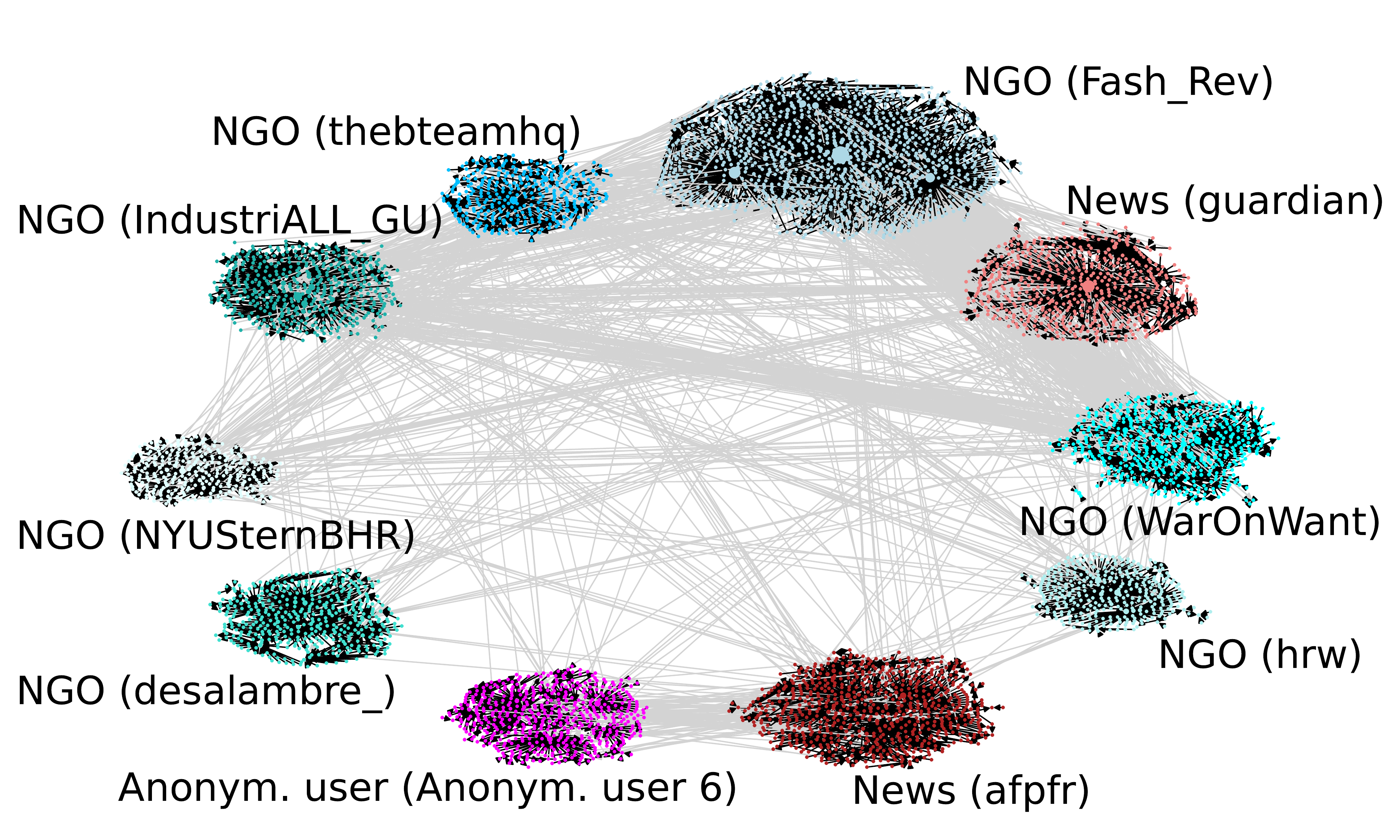}
			\tiny
			\raisebox{44pt}[0pt][0pt]{
				\begin{tabular}{c}
					\hline\hline
					Rank\\\hline
					1\\
					2\\
					3\\
					4\\
					5\\
					6\\
					7\\
					8\\
					9\\
					10\\\hline\hline
				\end{tabular}
				\vspace{5pt}
				\begin{tabular}{lc}
					\hline\hline
					\raisebox{1pt}[0pt][0pt]{\colorbox{white}{}} User & TC\\\hline
					\raisebox{1pt}[0pt][0pt]{\colorbox{lightblue}{}} Fash\_Rev & 150\\
					\raisebox{1pt}[0pt][0pt]{\colorbox{lightcoral}{}} guardian & 71\\
					\raisebox{1pt}[0pt][0pt]{\colorbox{lightblue}{}} BritishVogue & 62\\
					\raisebox{1pt}[0pt][0pt]{\colorbox{lightblue}{}} BoF & 40\\
					\raisebox{1pt}[0pt][0pt]{\colorbox{lightseagreen}{}} IndustriALL\_GU & 31\\
					\raisebox{1pt}[0pt][0pt]{\colorbox{deepskyblue}{}} thebteamhq & 26\\
					\raisebox{1pt}[0pt][0pt]{\colorbox{aqua}{}} WarOnWant & 25\\
					\raisebox{1pt}[0pt][0pt]{\colorbox{white}{}} Presa\_Diretta & 25\\
					\raisebox{1pt}[0pt][0pt]{\colorbox{white}{}} JasonMotlagh & 24\\
					\raisebox{1pt}[0pt][0pt]{\colorbox{white}{}} KooyJan & 22\\\hline\hline
				\end{tabular}
				\hfill
				\begin{tabular}{lc}
					\hline\hline
					\raisebox{1pt}[0pt][0pt]{\colorbox{white}{}} User & KC\\\hline
					\raisebox{1pt}[0pt][0pt]{\colorbox{lightblue}{}} Fash\_Rev & 26.5\\
					\raisebox{1pt}[0pt][0pt]{\colorbox{lightcoral}{}} guardian & 16.3\\
					\raisebox{1pt}[0pt][0pt]{\colorbox{lightblue}{}} BritishVogue & 14.3\\
					\raisebox{1pt}[0pt][0pt]{\colorbox{lightblue}{}} BoF & 10.1\\
					\raisebox{1pt}[0pt][0pt]{\colorbox{white}{}} Presa\_Diretta & 7.9\\
					\raisebox{1pt}[0pt][0pt]{\colorbox{deepskyblue}{}} thebteamhq & 7.8\\
					\raisebox{1pt}[0pt][0pt]{\colorbox{lightseagreen}{}} IndustriALL\_GU & 7.5\\
					\raisebox{1pt}[0pt][0pt]{\colorbox{aqua}{}} WarOnWant & 7.4\\
					\raisebox{1pt}[0pt][0pt]{\colorbox{white}{}} KooyJan & 7.4\\
					\raisebox{1pt}[0pt][0pt]{\colorbox{white}{}} AJEnglish & 6.5\\\hline\hline
				\end{tabular}
			}
			\normalsize
		\end{minipage}
	}

	\subfloat[Year $2015$]{
		\begin{minipage}{1.1\textwidth}
			\includegraphics[width=.4\textwidth]{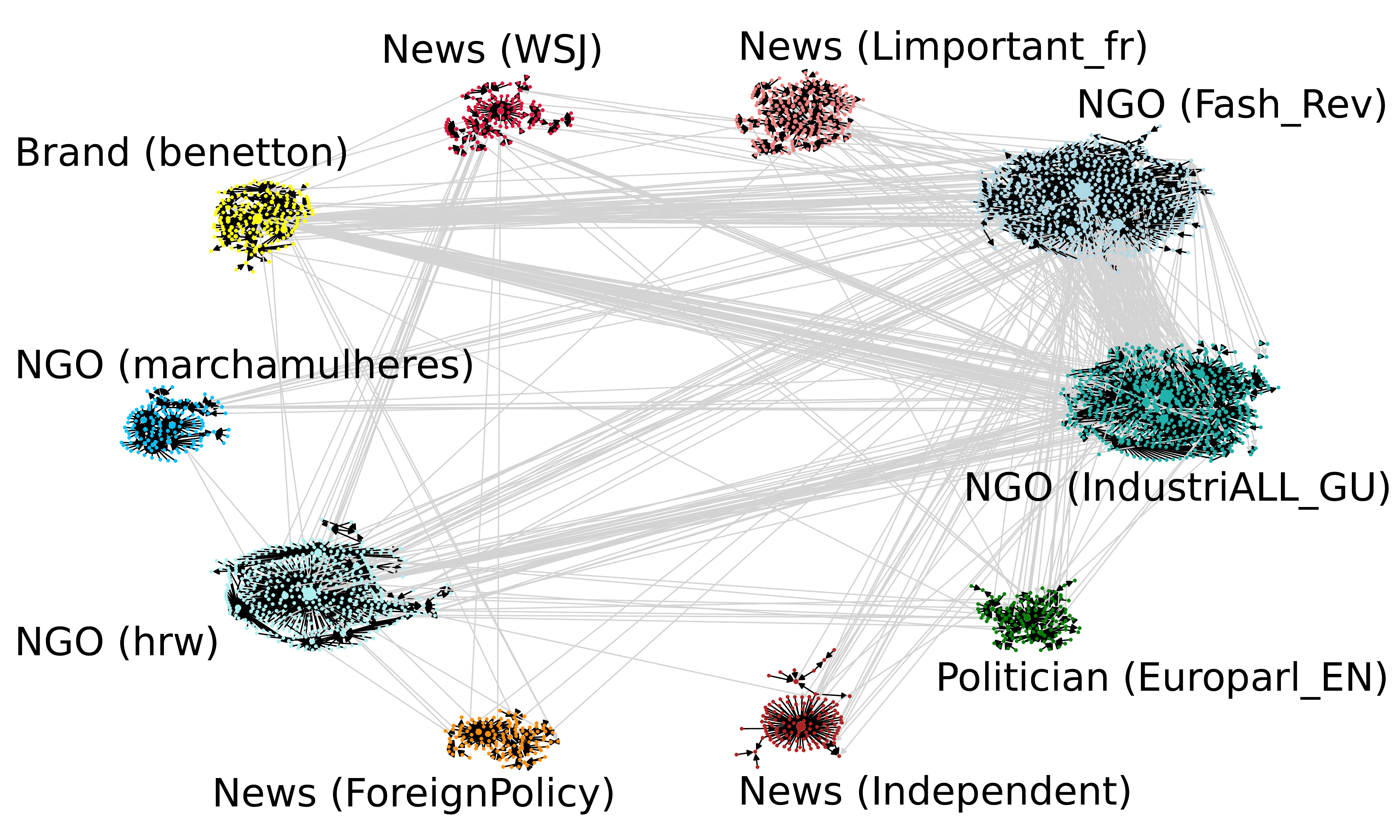}
			\tiny
			\raisebox{44pt}[0pt][0pt]{
				\begin{tabular}{c}
					\hline\hline
					Rank\\\hline
					1\\
					2\\
					3\\
					4\\
					5\\
					6\\
					7\\
					8\\
					9\\
					10\\\hline\hline
				\end{tabular}
				\vspace{5pt}
				\begin{tabular}{lc}
					\hline\hline
					\raisebox{1pt}[0pt][0pt]{\colorbox{white}{}} User & TC\\\hline
					\raisebox{1pt}[0pt][0pt]{\colorbox{lightblue}{}} Fash\_Rev & 131\\
					\raisebox{1pt}[0pt][0pt]{\colorbox{paleturquoise}{}} hrw & 85\\
					\raisebox{1pt}[0pt][0pt]{\colorbox{lightseagreen}{}} IndustriALL\_GU & 80\\
					\raisebox{1pt}[0pt][0pt]{\colorbox{lightblue}{}} guardian & 69\\
					\raisebox{1pt}[0pt][0pt]{\colorbox{yellow}{}} benetton & 44\\
					\raisebox{1pt}[0pt][0pt]{\colorbox{lightblue}{}} labourlabel & 44\\
					\raisebox{1pt}[0pt][0pt]{\colorbox{lightseagreen}{}} ILRF & 40\\
					\raisebox{1pt}[0pt][0pt]{\colorbox{firebrick}{}} Independent & 38\\
					\raisebox{1pt}[0pt][0pt]{\colorbox{lightseagreen}{}} equaltimes & 37\\
					\raisebox{1pt}[0pt][0pt]{\colorbox{lightseagreen}{}} cleanclothes & 33\\\hline\hline
				\end{tabular}
				\vspace{5pt}
				\begin{tabular}{lc}
					\hline\hline
					\raisebox{1pt}[0pt][0pt]{\colorbox{white}{}} User & KC\\\hline
					\raisebox{1pt}[0pt][0pt]{\colorbox{lightblue}{}} Fash\_Rev & 24.1\\
					\raisebox{1pt}[0pt][0pt]{\colorbox{paleturquoise}{}} hrw & 19.6\\
					\raisebox{1pt}[0pt][0pt]{\colorbox{lightblue}{}} guardian & 15.3\\
					\raisebox{1pt}[0pt][0pt]{\colorbox{lightseagreen}{}} IndustriALL\_GU & 14.7\\
					\raisebox{1pt}[0pt][0pt]{\colorbox{firebrick}{}} Independent & 10.1\\
					\raisebox{1pt}[0pt][0pt]{\colorbox{yellow}{}} benetton & 10.0\\
					\raisebox{1pt}[0pt][0pt]{\colorbox{lightseagreen}{}} ILRF & 9.7\\
					\raisebox{1pt}[0pt][0pt]{\colorbox{lightblue}{}} labourlabel & 9.1\\
					\raisebox{1pt}[0pt][0pt]{\colorbox{lightseagreen}{}} equaltimes & 9.0\\
					\raisebox{1pt}[0pt][0pt]{\colorbox{lightseagreen}{}} cleanclothes & 8.7\\\hline\hline
				\end{tabular}
			}
			\normalsize
		\end{minipage}
	}
	
	\subfloat[Year $2016$]{
		\begin{minipage}{1.1\textwidth}
			\includegraphics[width=.4\textwidth]{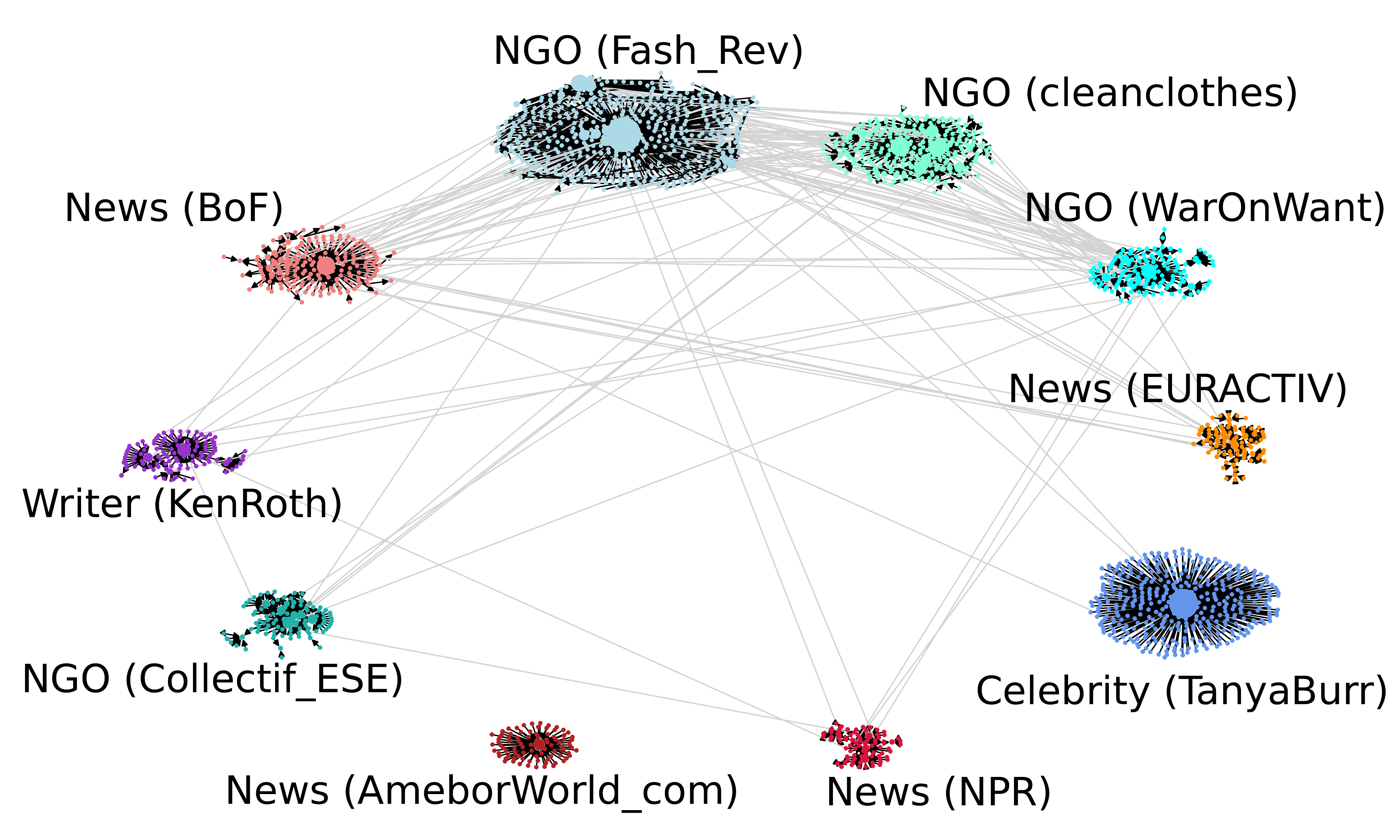}
			\tiny
			\raisebox{44pt}[0pt][0pt]{
				\begin{tabular}{c}
					\hline\hline
					Rank\\\hline
					1\\
					2\\
					3\\
					4\\
					5\\
					6\\
					7\\
					8\\
					9\\
					10\\\hline\hline
				\end{tabular}
				\vspace{5pt}
				\begin{tabular}{lc}
					\hline\hline
					\raisebox{1pt}[0pt][0pt]{\colorbox{white}{}} User & TC\\\hline
					\raisebox{1pt}[0pt][0pt]{\colorbox{lightblue}{}} Fash\_Rev & 665\\
					\raisebox{1pt}[0pt][0pt]{\colorbox{cornflowerblue}{}} TanyaBurr & 447\\
					\raisebox{1pt}[0pt][0pt]{\colorbox{aquamarine}{}} cleanclothes & 187\\
					\raisebox{1pt}[0pt][0pt]{\colorbox{lightcoral}{}} BoF & 169\\
					\raisebox{1pt}[0pt][0pt]{\colorbox{aquamarine}{}} ILRF & 150\\
					\raisebox{1pt}[0pt][0pt]{\colorbox{lightblue}{}} IndustriALL\_GU & 149\\
					\raisebox{1pt}[0pt][0pt]{\colorbox{aqua}{}} WarOnWant & 108\\
					\raisebox{1pt}[0pt][0pt]{\colorbox{darkorchid}{}} KenRoth & 81\\
					\raisebox{1pt}[0pt][0pt]{\colorbox{lightblue}{}} liviafirth & 79\\
					\raisebox{1pt}[0pt][0pt]{\colorbox{aquamarine}{}} labourlabel & 73\\\hline\hline
				\end{tabular}
				\hfill
				\begin{tabular}{lc}
					\hline\hline
					\raisebox{1pt}[0pt][0pt]{\colorbox{white}{}} User & KC\\\hline
					\raisebox{1pt}[0pt][0pt]{\colorbox{lightblue}{}} Fash\_Rev & 117.1\\
					\raisebox{1pt}[0pt][0pt]{\colorbox{cornflowerblue}{}} TanyaBurr & 81.9\\
					\raisebox{1pt}[0pt][0pt]{\colorbox{lightcoral}{}} BoF & 31.7\\
					\raisebox{1pt}[0pt][0pt]{\colorbox{aquamarine}{}} cleanclothes & 31.5\\
					\raisebox{1pt}[0pt][0pt]{\colorbox{aquamarine}{}} ILRF & 23.1\\
					\raisebox{1pt}[0pt][0pt]{\colorbox{lightblue}{}} IndustriALL\_GU & 22.7\\
					\raisebox{1pt}[0pt][0pt]{\colorbox{aqua}{}} WarOnWant & 20.5\\
					\raisebox{1pt}[0pt][0pt]{\colorbox{darkorchid}{}} KenRoth & 16.9\\
					\raisebox{1pt}[0pt][0pt]{\colorbox{white}{}} afpfr & 15.3\\
					\raisebox{1pt}[0pt][0pt]{\colorbox{aquamarine}{}} labourlabel & 14.2\\\hline\hline
				\end{tabular}
			}
			\normalsize
		\end{minipage}
	}
	
	\subfloat[Year $2017$]{
		\begin{minipage}{1.1\textwidth}
			\includegraphics[width=.4\textwidth]{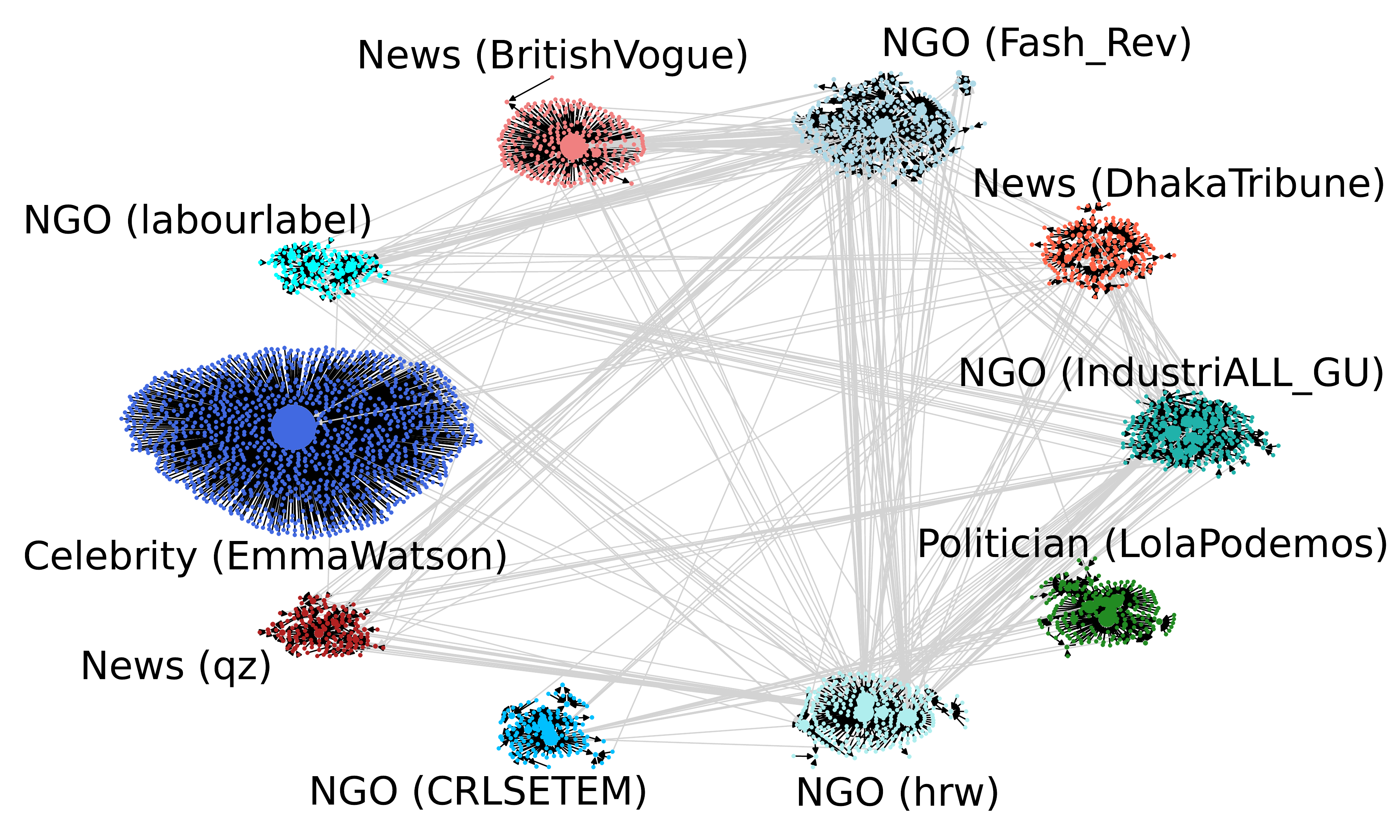}
			\tiny
			\raisebox{44pt}[0pt][0pt]{
				\begin{tabular}{c}
					\hline\hline
					Rank\\\hline
					1\\
					2\\
					3\\
					4\\
					5\\
					6\\
					7\\
					8\\
					9\\
					10\\\hline\hline
				\end{tabular}
				\vspace{5pt}
				\begin{tabular}{lc}
					\hline\hline
					\raisebox{1pt}[0pt][0pt]{\colorbox{white}{}} User & TC\\\hline
					\raisebox{1pt}[0pt][0pt]{\colorbox{royalblue}{}} EmmaWatson & 1\,151\\
					\raisebox{1pt}[0pt][0pt]{\colorbox{lightcoral}{}} BritishVogue & 360\\
					\raisebox{1pt}[0pt][0pt]{\colorbox{lightblue}{}} Fash\_Rev & 198\\
					\raisebox{1pt}[0pt][0pt]{\colorbox{paleturquoise}{}} hrw & 188\\
					\raisebox{1pt}[0pt][0pt]{\colorbox{paleturquoise}{}} ajkashy & 166\\
					\raisebox{1pt}[0pt][0pt]{\colorbox{forestgreen}{}} LolaPodemos & 154\\
					\raisebox{1pt}[0pt][0pt]{\colorbox{paleturquoise}{}} cleanclothes & 153\\
					\raisebox{1pt}[0pt][0pt]{\colorbox{lightseagreen}{}} IndustriALL\_GU & 124\\
					\raisebox{1pt}[0pt][0pt]{\colorbox{deepskyblue}{}} CRLSETEM & 113\\
					\raisebox{1pt}[0pt][0pt]{\colorbox{forestgreen}{}} carnecrudaradio & 95\\\hline\hline
				\end{tabular}
				\hfill
				\begin{tabular}{lc}
					\hline\hline
					\raisebox{1pt}[0pt][0pt]{\colorbox{white}{}} User & KC\\\hline
					\raisebox{1pt}[0pt][0pt]{\colorbox{royalblue}{}} EmmaWatson & 216.9\\
					\raisebox{1pt}[0pt][0pt]{\colorbox{lightcoral}{}} BritishVogue & 64.1\\
					\raisebox{1pt}[0pt][0pt]{\colorbox{lightblue}{}} Fash\_Rev & 35.1\\
					\raisebox{1pt}[0pt][0pt]{\colorbox{paleturquoise}{}} hrw & 32.7\\
					\raisebox{1pt}[0pt][0pt]{\colorbox{paleturquoise}{}} cleanclothes & 27.9\\
					\raisebox{1pt}[0pt][0pt]{\colorbox{paleturquoise}{}} ajkashy & 26.3\\
					\raisebox{1pt}[0pt][0pt]{\colorbox{forestgreen}{}} LolaPodemos & 26.2\\
					\raisebox{1pt}[0pt][0pt]{\colorbox{deepskyblue}{}} CRLSETEM & 21.4\\
					\raisebox{1pt}[0pt][0pt]{\colorbox{lightseagreen}{}} IndustriALL\_GU & 20.7\\
					\raisebox{1pt}[0pt][0pt]{\colorbox{forestgreen}{}} carnecrudaradio & 17.1\\\hline\hline
				\end{tabular}
			}
			\normalsize
		\end{minipage}
	}
	
	\vspace{10pt}
	\begin{minipage}{1.1\textwidth}
		\begin{center}
			\fbox{%
				\parbox{.65\textwidth}{
					\centering
					\small
					\begin{tabular}{lllll}
						\raisebox{2pt}[0pt][0pt]{\colorbox{white}{}}\raisebox{2pt}[0pt][0pt]{\colorbox{white}{}}\raisebox{2pt}[0pt][0pt]{\colorbox{white}{}}\raisebox{2pt}[0pt][0pt]{\colorbox{white}{}}\raisebox{2pt}[0pt][0pt]{\colorbox{white}{}}\raisebox{2pt}[0pt][0pt]{\colorbox{white}{}}\raisebox{2pt}[0pt][0pt]{\colorbox{white}{}}\raisebox{2pt}[0pt][0pt]{\colorbox{white}{}} & Activist && \raisebox{2pt}[0pt][0pt]{\colorbox{darkgreen}{}}\raisebox{2pt}[0pt][0pt]{\colorbox{green}{}}\raisebox{2pt}[0pt][0pt]{\colorbox{forestgreen}{}}\raisebox{2pt}[0pt][0pt]{\colorbox{white}{}}\raisebox{2pt}[0pt][0pt]{\colorbox{white}{}} & Politician\\
						\raisebox{2pt}[0pt][0pt]{\colorbox{paleturquoise}{}}\raisebox{2pt}[0pt][0pt]{\colorbox{lightseagreen}{}}\raisebox{2pt}[0pt][0pt]{\colorbox{deepskyblue}{}}\raisebox{2pt}[0pt][0pt]{\colorbox{lightblue}{}}\raisebox{2pt}[0pt][0pt]{\colorbox{turquoise}{}}\raisebox{2pt}[0pt][0pt]{\colorbox{aqua}{}}\raisebox{2pt}[0pt][0pt]{\colorbox{aquamarine}{}}\raisebox{2pt}[0pt][0pt]{\colorbox{lightcyan}{}} & \hspace{5pt}- NGO && \raisebox{2pt}[0pt][0pt]{\colorbox{firebrick}{}}\raisebox{2pt}[0pt][0pt]{\colorbox{crimson}{}}\raisebox{2pt}[0pt][0pt]{\colorbox{tomato}{}}\raisebox{2pt}[0pt][0pt]{\colorbox{lightcoral}{}}\raisebox{2pt}[0pt][0pt]{\colorbox{darkorange}{}} & News agency\\
						\raisebox{2pt}[0pt][0pt]{\colorbox{blueviolet}{}}\raisebox{2pt}[0pt][0pt]{\colorbox{darkorchid}{}}\raisebox{2pt}[0pt][0pt]{\colorbox{mediumorchid}{}}\raisebox{2pt}[0pt][0pt]{\colorbox{white}{}}\raisebox{2pt}[0pt][0pt]{\colorbox{white}{}}\raisebox{2pt}[0pt][0pt]{\colorbox{white}{}}\raisebox{2pt}[0pt][0pt]{\colorbox{white}{}}\raisebox{2pt}[0pt][0pt]{\colorbox{white}{}} & \hspace{5pt}- Writer && \raisebox{2pt}[0pt][0pt]{\colorbox{fuchsia}{}}\raisebox{2pt}[0pt][0pt]{\colorbox{violet}{}}\raisebox{2pt}[0pt][0pt]{\colorbox{white}{}}\raisebox{2pt}[0pt][0pt]{\colorbox{white}{}}\raisebox{2pt}[0pt][0pt]{\colorbox{white}{}} & Anonymous user\\
						\raisebox{2pt}[0pt][0pt]{\colorbox{royalblue}{}}\raisebox{2pt}[0pt][0pt]{\colorbox{cornflowerblue}{}}\raisebox{2pt}[0pt][0pt]{\colorbox{white}{}}\raisebox{2pt}[0pt][0pt]{\colorbox{white}{}}\raisebox{2pt}[0pt][0pt]{\colorbox{white}{}}\raisebox{2pt}[0pt][0pt]{\colorbox{white}{}}\raisebox{2pt}[0pt][0pt]{\colorbox{white}{}}\raisebox{2pt}[0pt][0pt]{\colorbox{white}{}} & \hspace{5pt}- Celebrity && \raisebox{2pt}[0pt][0pt]{\colorbox{yellow}{}}\raisebox{2pt}[0pt][0pt]{\colorbox{white}{}}\raisebox{2pt}[0pt][0pt]{\colorbox{white}{}}\raisebox{2pt}[0pt][0pt]{\colorbox{white}{}}\raisebox{2pt}[0pt][0pt]{\colorbox{white}{}} & Brand
					\end{tabular}
				}
			}
		\end{center}
	\end{minipage}
	\caption{}
\end{figure}

\begin{figure}
	\thisfloatpagestyle{empty}
	\ContinuedFloat
	\subfloat[Year $2018$]{
		\begin{minipage}{1.1\textwidth}
			\includegraphics[width=.4\textwidth]{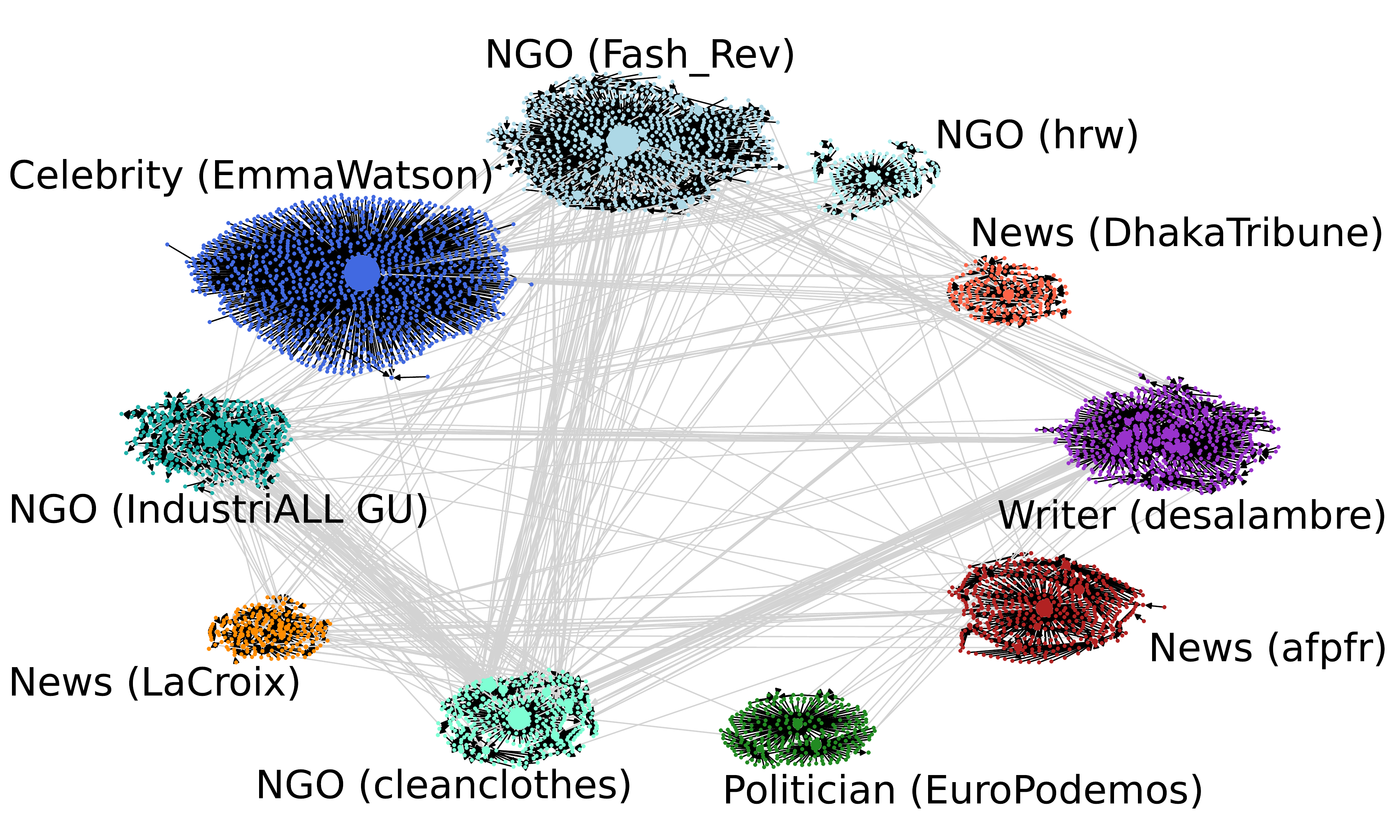}
			\tiny
			\raisebox{44pt}[0pt][0pt]{
				\begin{tabular}{c}
					\hline\hline
					Rank\\\hline
					1\\
					2\\
					3\\
					4\\
					5\\
					6\\
					7\\
					8\\
					9\\
					10\\\hline\hline
				\end{tabular}
				\vspace{5pt}
				\begin{tabular}{lc}
					\hline\hline
					\raisebox{1pt}[0pt][0pt]{\colorbox{white}{}} User & TC\\\hline
					\raisebox{1pt}[0pt][0pt]{\colorbox{royalblue}{}} EmmaWatson & 717\\
					\raisebox{1pt}[0pt][0pt]{\colorbox{lightblue}{}} Fash\_Rev & 535\\
					\raisebox{1pt}[0pt][0pt]{\colorbox{aquamarine}{}} cleanclothes & 258\\
					\raisebox{1pt}[0pt][0pt]{\colorbox{firebrick}{}} afpfr & 150\\
					\raisebox{1pt}[0pt][0pt]{\colorbox{lightseagreen}{}} IndustriALL\_GU & 128\\
					\raisebox{1pt}[0pt][0pt]{\colorbox{aquamarine}{}} ethicaltrade & 101\\
					\raisebox{1pt}[0pt][0pt]{\colorbox{lightseagreen}{}} uniglobalunion & 93\\
					\raisebox{1pt}[0pt][0pt]{\colorbox{darkorchid}{}} desalambre & 86\\
					\raisebox{1pt}[0pt][0pt]{\colorbox{white}{}} fashionsnap & 83\\
					\raisebox{1pt}[0pt][0pt]{\colorbox{darkorchid}{}} CRLSETEM & 81\\\hline\hline
				\end{tabular}
				\hfill
				\begin{tabular}{lc}
					\hline\hline
					\raisebox{1pt}[0pt][0pt]{\colorbox{white}{}} User & KC\\\hline
					\raisebox{1pt}[0pt][0pt]{\colorbox{royalblue}{}} EmmaWatson & 144.4\\
					\raisebox{1pt}[0pt][0pt]{\colorbox{lightblue}{}} Fash\_Rev & 94.5\\
					\raisebox{1pt}[0pt][0pt]{\colorbox{aquamarine}{}} cleanclothes & 45.1\\
					\raisebox{1pt}[0pt][0pt]{\colorbox{firebrick}{}} afpfr & 31.9\\
					\raisebox{1pt}[0pt][0pt]{\colorbox{lightseagreen}{}} IndustriALL\_GU & 23.2\\
					\raisebox{1pt}[0pt][0pt]{\colorbox{white}{}} fashionsnap & 18.9\\
					\raisebox{1pt}[0pt][0pt]{\colorbox{darkorchid}{}} desalambre & 18.2\\
					\raisebox{1pt}[0pt][0pt]{\colorbox{paleturquoise}{}} hrw & 17.0\\
					\raisebox{1pt}[0pt][0pt]{\colorbox{lightseagreen}{}} uniglobalunion & 16.3\\
					\raisebox{1pt}[0pt][0pt]{\colorbox{darkorchid}{}} CRLSETEM & 16.1\\\hline\hline
				\end{tabular}
			}
			\normalsize
		\end{minipage}
	}
	
	\subfloat[Year $2019$]{
		\begin{minipage}{1.1\textwidth}
			\includegraphics[width=.4\textwidth]{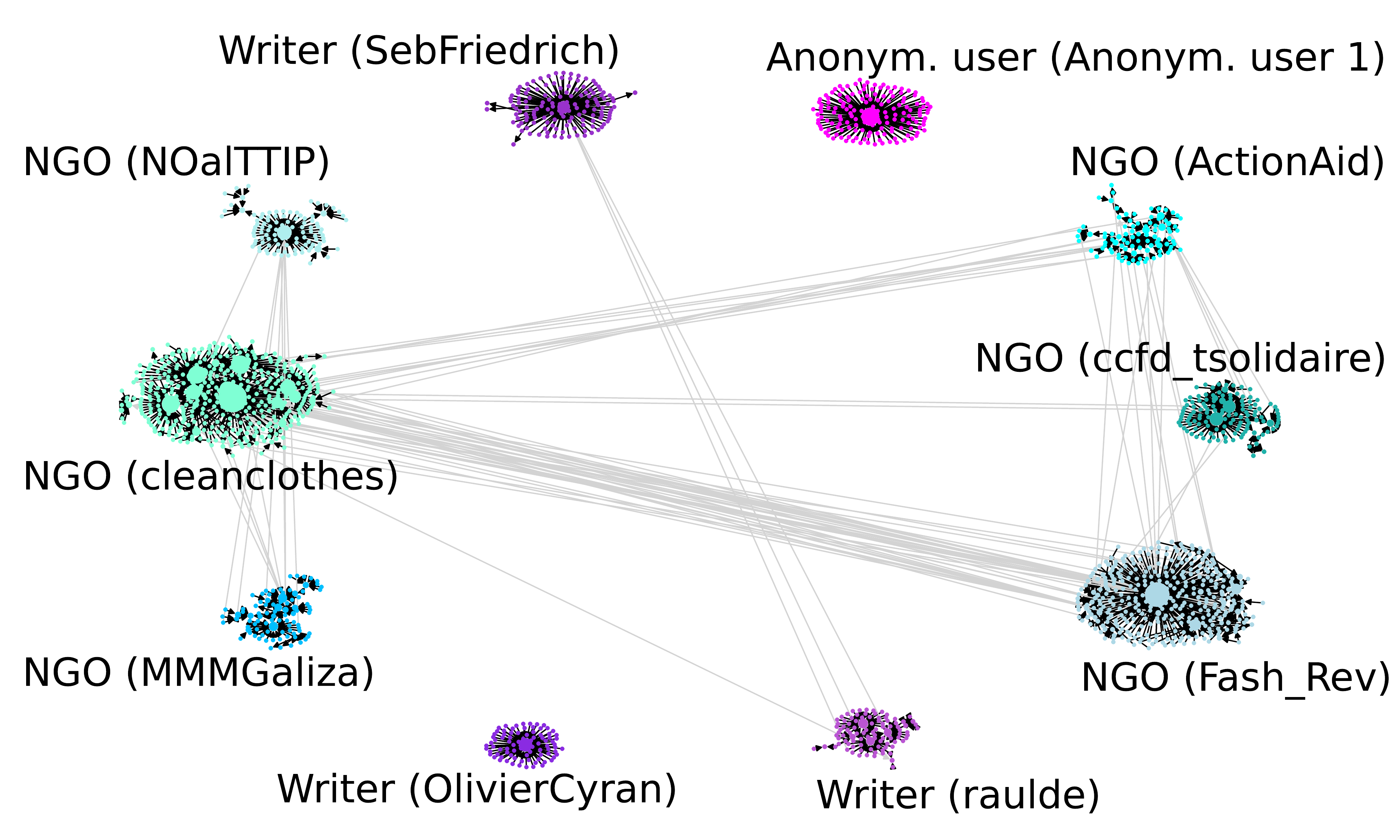}
			\tiny
			\raisebox{44pt}[0pt][0pt]{
				\begin{tabular}{c}
					\hline\hline
					Rank\\\hline
					1\\
					2\\
					3\\
					4\\
					5\\
					6\\
					7\\
					8\\
					9\\
					10\\\hline\hline
				\end{tabular}
				\vspace{5pt}
				\begin{tabular}{lc}
					\hline\hline
					\raisebox{1pt}[0pt][0pt]{\colorbox{white}{}} User & TC\\\hline
					\raisebox{1pt}[0pt][0pt]{\colorbox{aquamarine}{}} cleanclothes & 436\\
					\raisebox{1pt}[0pt][0pt]{\colorbox{lightblue}{}} Fash\_Rev & 318\\
					\raisebox{1pt}[0pt][0pt]{\colorbox{aquamarine}{}} SolidarityCntr & 178\\
					\raisebox{1pt}[0pt][0pt]{\colorbox{fuchsia}{}} Anonym.\ user 1 & 161\\
					\raisebox{1pt}[0pt][0pt]{\colorbox{aquamarine}{}} ILRF & 160\\
					\raisebox{1pt}[0pt][0pt]{\colorbox{aquamarine}{}} TansyHoskins & 154\\
					\raisebox{1pt}[0pt][0pt]{\colorbox{aquamarine}{}} uniglobalunion & 121\\
					\raisebox{1pt}[0pt][0pt]{\colorbox{aquamarine}{}} IndustriALL\_GU & 117\\
					\raisebox{1pt}[0pt][0pt]{\colorbox{aquamarine}{}} 4WorkerRights & 111\\
					\raisebox{1pt}[0pt][0pt]{\colorbox{paleturquoise}{}} NOalTTIP & 104\\\hline\hline
				\end{tabular}
				\hfill
				\begin{tabular}{lc}
					\hline\hline
					\raisebox{1pt}[0pt][0pt]{\colorbox{white}{}} User & KC\\\hline
					\raisebox{1pt}[0pt][0pt]{\colorbox{aquamarine}{}} cleanclothes & 71.6\\
					\raisebox{1pt}[0pt][0pt]{\colorbox{lightblue}{}} Fash\_Rev & 59.3\\
					\raisebox{1pt}[0pt][0pt]{\colorbox{fuchsia}{}} Anonym.\ user 1 & 31.6\\
					\raisebox{1pt}[0pt][0pt]{\colorbox{aquamarine}{}} TansyHoskins & 26.3\\
					\raisebox{1pt}[0pt][0pt]{\colorbox{aquamarine}{}} SolidarityCntr & 25.0\\
					\raisebox{1pt}[0pt][0pt]{\colorbox{aquamarine}{}} ILRF & 24.6\\
					\raisebox{1pt}[0pt][0pt]{\colorbox{paleturquoise}{}} NOalTTIP & 21.3\\
					\raisebox{1pt}[0pt][0pt]{\colorbox{aquamarine}{}} uniglobalunion & 19.2\\
					\raisebox{1pt}[0pt][0pt]{\colorbox{blueviolet}{}} OlivierCyran & 18.9\\
					\raisebox{1pt}[0pt][0pt]{\colorbox{lightseagreen}{}} ccfd\_tsolidaire & 17.6\\\hline\hline
				\end{tabular}
			}
			\normalsize
		\end{minipage}
	}
	
	\subfloat[Year $2020$]{
		\begin{minipage}{1.1\textwidth}
			\includegraphics[width=.4\textwidth]{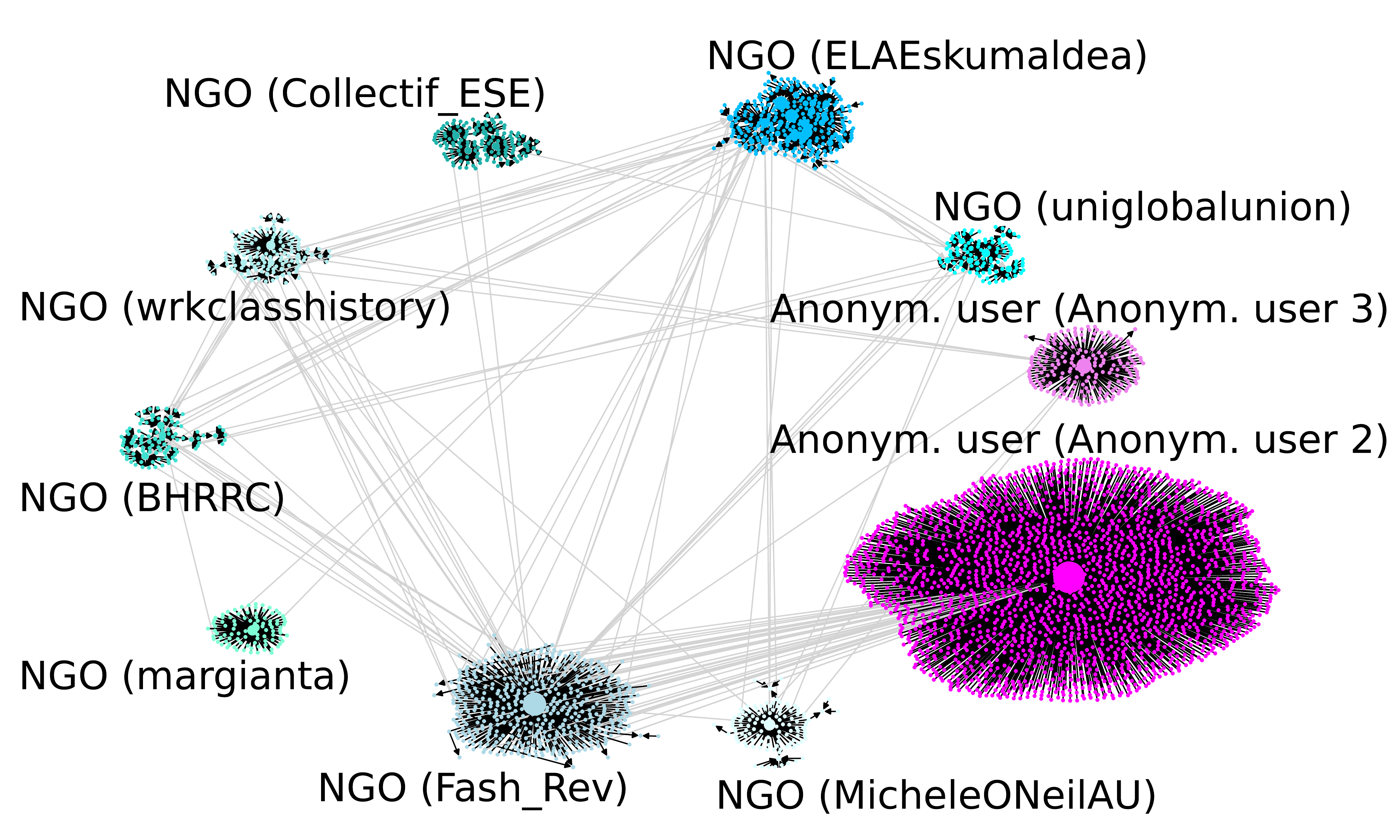}
			\tiny
			\raisebox{44pt}[0pt][0pt]{
				\begin{tabular}{c}
					\hline\hline
					Rank\\\hline
					1\\
					2\\
					3\\
					4\\
					5\\
					6\\
					7\\
					8\\
					9\\
					10\\\hline\hline
				\end{tabular}
				\vspace{5pt}
				\begin{tabular}{lc}
					\hline\hline
					\raisebox{1pt}[0pt][0pt]{\colorbox{white}{}} User & TC\\\hline
					\raisebox{1pt}[0pt][0pt]{\colorbox{fuchsia}{}} Anonym.\ user 2 & 540\\
					\raisebox{1pt}[0pt][0pt]{\colorbox{lightblue}{}} Fash\_Rev & 282\\
					\raisebox{1pt}[0pt][0pt]{\colorbox{violet}{}} Anonym.\ user 3 & 112\\
					\raisebox{1pt}[0pt][0pt]{\colorbox{deepskyblue}{}} ELAEskumaldea & 93\\
					\raisebox{1pt}[0pt][0pt]{\colorbox{deepskyblue}{}} emakumemartxa & 83\\
					\raisebox{1pt}[0pt][0pt]{\colorbox{white}{}} Awaj\_fdn & 63\\
					\raisebox{1pt}[0pt][0pt]{\colorbox{lightcyan}{}} MicheleONeilAU & 60\\
					\raisebox{1pt}[0pt][0pt]{\colorbox{deepskyblue}{}} CRLSETEM & 59\\
					\raisebox{1pt}[0pt][0pt]{\colorbox{aquamarine}{}} margianta & 51\\
					\raisebox{1pt}[0pt][0pt]{\colorbox{white}{}} Independent&  50\\\hline\hline
				\end{tabular}
				\hfill
				\begin{tabular}{lc}
					\hline\hline
					\raisebox{1pt}[0pt][0pt]{\colorbox{white}{}} User & KC\\\hline
					\raisebox{1pt}[0pt][0pt]{\colorbox{fuchsia}{}} Anonym.\ user 2 & 110.9\\
					\raisebox{1pt}[0pt][0pt]{\colorbox{lightblue}{}} Fash\_Rev & 58.0\\
					\raisebox{1pt}[0pt][0pt]{\colorbox{violet}{}} Anonym.\ user 3 & 24.9\\
					\raisebox{1pt}[0pt][0pt]{\colorbox{deepskyblue}{}} ELAEskumaldea & 17.6\\
					\raisebox{1pt}[0pt][0pt]{\colorbox{deepskyblue}{}} emakumemartxa & 16.2\\
					\raisebox{1pt}[0pt][0pt]{\colorbox{lightcyan}{}} MicheleONeilAU & 14.3\\
					\raisebox{1pt}[0pt][0pt]{\colorbox{white}{}} Awaj\_fdn & 13.2\\
					\raisebox{1pt}[0pt][0pt]{\colorbox{aquamarine}{}} margianta & 12.7\\
					\raisebox{1pt}[0pt][0pt]{\colorbox{deepskyblue}{}} CRLSETEM & 11.0\\
					\raisebox{1pt}[0pt][0pt]{\colorbox{deepskyblue}{}} cleanclothes & 9.8\\\hline\hline
				\end{tabular}
			}
			\normalsize
		\end{minipage}
	}
	
	\subfloat[Year $2021$]{
		\begin{minipage}{1.1\textwidth}
			\includegraphics[width=.4\textwidth]{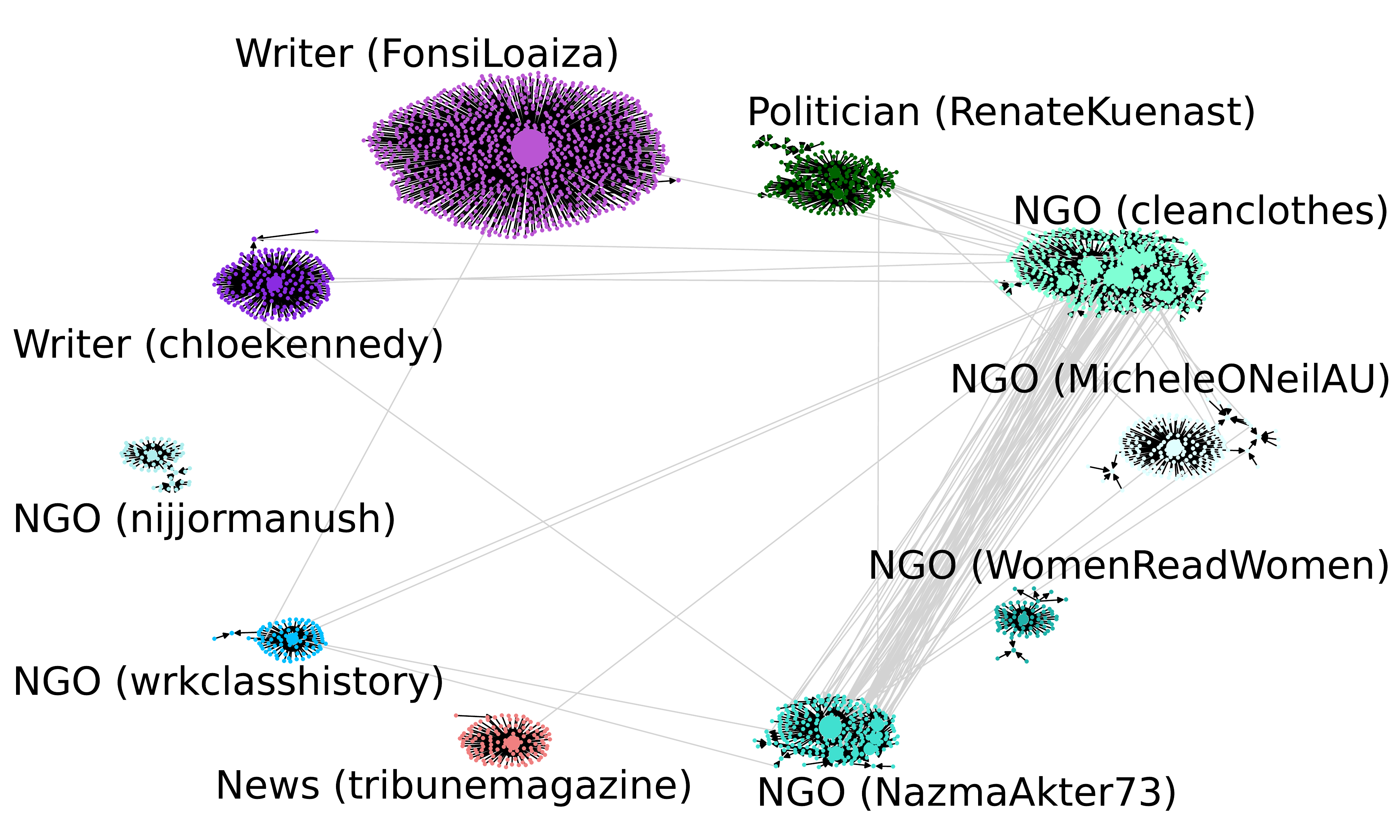}
			\tiny
			\raisebox{44pt}[0pt][0pt]{
				\begin{tabular}{c}
					\hline\hline
					Rank\\\hline
					1\\
					2\\
					3\\
					4\\
					5\\
					6\\
					7\\
					8\\
					9\\
					10\\\hline\hline
				\end{tabular}
				\vspace{5pt}
				\begin{tabular}{lc}
					\hline\hline
					\raisebox{1pt}[0pt][0pt]{\colorbox{white}{}} User & TC\\\hline
					\raisebox{1pt}[0pt][0pt]{\colorbox{mediumorchid}{}} FonsiLoaiza & 809\\
					\raisebox{1pt}[0pt][0pt]{\colorbox{turquoise}{}} NazmaAkter73 & 283\\
					\raisebox{1pt}[0pt][0pt]{\colorbox{aquamarine}{}} cleanclothes & 270\\
					\raisebox{1pt}[0pt][0pt]{\colorbox{aquamarine}{}} Fash\_Rev & 211\\
					\raisebox{1pt}[0pt][0pt]{\colorbox{aquamarine}{}} labourlabel & 192\\
					\raisebox{1pt}[0pt][0pt]{\colorbox{lightcyan}{}} MicheleONeilAU & 133\\
					\raisebox{1pt}[0pt][0pt]{\colorbox{aquamarine}{}} voguemagazine & 106\\
					\raisebox{1pt}[0pt][0pt]{\colorbox{turquoise}{}} IndustriALL\_GU & 103\\
					\raisebox{1pt}[0pt][0pt]{\colorbox{lightcoral}{}} tribunemagazine & 102\\
					\raisebox{1pt}[0pt][0pt]{\colorbox{aquamarine}{}} TansyHoskins & 100\\\hline\hline
				\end{tabular}
				\hfill
				\begin{tabular}{lc}
					\hline\hline
					\raisebox{1pt}[0pt][0pt]{\colorbox{white}{}} User & KC\\\hline
					\raisebox{1pt}[0pt][0pt]{\colorbox{mediumorchid}{}} FonsiLoaiza & 155.6\\
					\raisebox{1pt}[0pt][0pt]{\colorbox{turquoise}{}} NazmaAkter73 & 51.1\\
					\raisebox{1pt}[0pt][0pt]{\colorbox{aquamarine}{}} cleanclothes & 44.0\\
					\raisebox{1pt}[0pt][0pt]{\colorbox{aquamarine}{}} Fash\_Rev & 34.6\\
					\raisebox{1pt}[0pt][0pt]{\colorbox{aquamarine}{}} labourlabel & 27.7\\
					\raisebox{1pt}[0pt][0pt]{\colorbox{lightcyan}{}} MicheleONeilAU & 27.6\\
					\raisebox{1pt}[0pt][0pt]{\colorbox{lightcoral}{}} tribunemagazine & 21.8\\
					\raisebox{1pt}[0pt][0pt]{\colorbox{aquamarine}{}} voguemagazine & 19.1\\
					\raisebox{1pt}[0pt][0pt]{\colorbox{blueviolet}{}} leoinlaurent & 18.0\\
					\raisebox{1pt}[0pt][0pt]{\colorbox{blueviolet}{}} chIoekennedy & 18.0\\\hline\hline
				\end{tabular}
			}
			\normalsize
		\end{minipage}
	}
	
	\subfloat[Year $2022$]{
		\begin{minipage}{1.1\textwidth}
			\includegraphics[width=.38\textwidth]{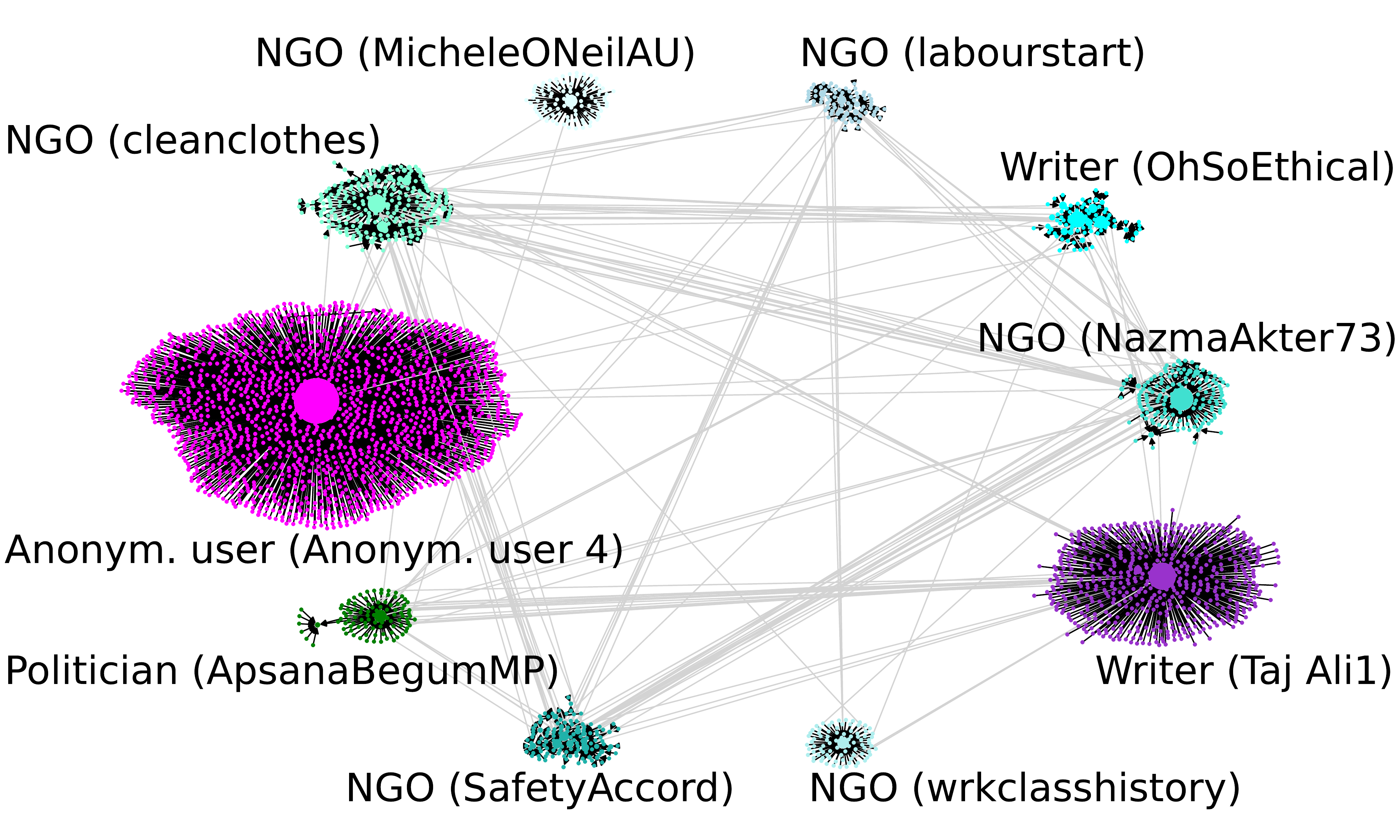}
			\tiny
			\raisebox{44pt}[0pt][0pt]{
				\begin{tabular}{c}
					\hline\hline
					Rank\\\hline
					1\\
					2\\
					3\\
					4\\
					5\\
					6\\
					7\\
					8\\
					9\\
					10\\\hline\hline
				\end{tabular}
				\vspace{5pt}
				\begin{tabular}{lc}
					\hline\hline
					\raisebox{1pt}[0pt][0pt]{\colorbox{white}{}} User & TC\\\hline
					\raisebox{1pt}[0pt][0pt]{\colorbox{fuchsia}{}} Anonym.\ user 4 & 1\,159\\
					\raisebox{1pt}[0pt][0pt]{\colorbox{darkorchid}{}} Taj\_Ali1 & 402\\
					\raisebox{1pt}[0pt][0pt]{\colorbox{turquoise}{}} NazmaAkter73 & 279\\
					\raisebox{1pt}[0pt][0pt]{\colorbox{aquamarine}{}} cleanclothes & 165\\
					\raisebox{1pt}[0pt][0pt]{\colorbox{aqua}{}} OhSoEthical & 135\\
					\raisebox{1pt}[0pt][0pt]{\colorbox{aqua}{}} nijjormanush & 82\\
					\raisebox{1pt}[0pt][0pt]{\colorbox{green}{}} ApsanaBegumMP & 77\\
					\raisebox{1pt}[0pt][0pt]{\colorbox{lightcyan}{}} MicheleONeilAU & 75\\
					\raisebox{1pt}[0pt][0pt]{\colorbox{paleturquoise}{}} wrkclasshistory & 63\\
					\raisebox{1pt}[0pt][0pt]{\colorbox{aquamarine}{}} Fash\_Rev & 55\\\hline\hline
				\end{tabular}
				\hfill
				\begin{tabular}{lc}
					\hline\hline
					\raisebox{1pt}[0pt][0pt]{\colorbox{white}{}} User & KC\\\hline
					\raisebox{1pt}[0pt][0pt]{\colorbox{fuchsia}{}} Anonym.\ user 4 & 225.6\\
					\raisebox{1pt}[0pt][0pt]{\colorbox{darkorchid}{}} Taj\_Ali1 & 79.5\\
					\raisebox{1pt}[0pt][0pt]{\colorbox{turquoise}{}} NazmaAkter73 & 53.4\\
					\raisebox{1pt}[0pt][0pt]{\colorbox{aquamarine}{}} cleanclothes & 31.2\\
					\raisebox{1pt}[0pt][0pt]{\colorbox{aqua}{}} OhSoEthical & 21.7\\
					\raisebox{1pt}[0pt][0pt]{\colorbox{green}{}} ApsanaBegumMP & 17.0\\
					\raisebox{1pt}[0pt][0pt]{\colorbox{lightcyan}{}} MicheleONeilAU & 16.9\\
					\raisebox{1pt}[0pt][0pt]{\colorbox{paleturquoise}{}} wrkclasshistory & 14.6\\
					\raisebox{1pt}[0pt][0pt]{\colorbox{aqua}{}} nijjormanush & 13.8\\
					\raisebox{1pt}[0pt][0pt]{\colorbox{white}{}} vanillaemotions & 12.8\\\hline\hline
				\end{tabular}
			}
			\normalsize
		\end{minipage}
	}
	\caption{Results of the quantitative multilayer network analysis.
		The left plots show all nodes (users) belonging to the ten largest communities of the respective year.
		Directed edges (interactions via retweets, replies, or mentions) within communities are marked black and edges across communities are marked gray.
		Nodes sizes are proportional to the receiver total communicability (TC) centrality measure.
		Color codes in the plots indicate the actor group of the most central node in each community.
		Each community is labeled by the actor group of its most central node as well as the most central node's user name in round brackets.
		The tables show the top ten users of the respective year according to receiver total communicability (TC) and receiver Katz centrality (KC).
		Color codes in the tables identify the users with their community in the plot.
		Blank spots in the centrality tables indicate that the user does not belong to any of the ten largest communities.}\label{fig:centralities_communities}
\end{figure}

The multilayer Louvain method additionally provides a partition of each year's set of users into densely connected communities of strongly varying sizes.
For all years, we obtain a large number of communities (typically around $1\,000$).
However, only around $50$ of these communities consist of ten or more users.
As we encounter many central nodes in the largest communities, we focus our attention on the ten largest communities per year.
The attempt to identify recurring communities across the years proved difficult.
One reason for this is the heterogeneous user base across the years: with $62\,500$, the sum of overall unique users over the ten years was only slightly below $74\,000$, the number of unique users summed up for each year separately.
Exceptions of highly central recurring users actively forming the discourse across the years can be found in the centrality results.
Furthermore, the assignment of communities to user nationalities proved difficult indicating a significant degree of internationality in the discourse.

We illustrate the ten largest communities per year in the graphics in Figure~\ref{fig:centralities_communities}.
Edges within communities are marked black and edges across communities are marked gray.
The connectivity structure within large communities often consists of one or few highly central users being reacted to by many non-central users.
The amount of inter-community connectivity greatly varies across the years, with partial explanations being revealed by the qualitative content analysis in \Cref{sec:results_content}.
Furthermore, the color coding of the nodes stems from the user groups ``Activist'', ``Politician'', ``News agency'', ``Anonymous user'', and ``Brand'' identified in the qualitative analysis of the tweets of the ten most central users described in \Cref{sec:methods_content}.
We use different color shades per user group to allow the identification of each community with its most central user in the centrality results in the respective year.
In an attempt to disclose the cohesion of structural network properties and our contentual analysis, we assign each community with a color code corresponding to the actor group of the most central user of that community.
The communities' actor groups as well as the most central nodes' user names in round brackets are depicted in the plots.
The same color codes in the tables in Figure~\ref{fig:centralities_communities} encode the affiliation of the ten most central users to the ten largest communities.
Blank spots in the centrality tables indicate that the user does not belong to any of the ten largest communities.

The prevalence of the five actor groups in the ten largest communities undergoes a significant temporal evolution.
Unsurprisingly, the year $2013$ is dominated by news agencies spreading information on the collapse.
Subsequent years witness their diminution and the emergence of large communication sub-networks around NGOs with $2020$ marking the record of eight out of ten communities classified as NGO.
A noteworthy development in the years $2020$ and $2022$ is that anonymous, i.e., private Twitter users are able to generate enough interaction to gather the largest community of that year around them; a role that was previously filled by news agencies, NGOs, or celebrities.

\subsection{Content analysis}\label{sec:results_content}

\begin{figure}
	\begin{center}
		\includegraphics[width=0.9\textwidth]{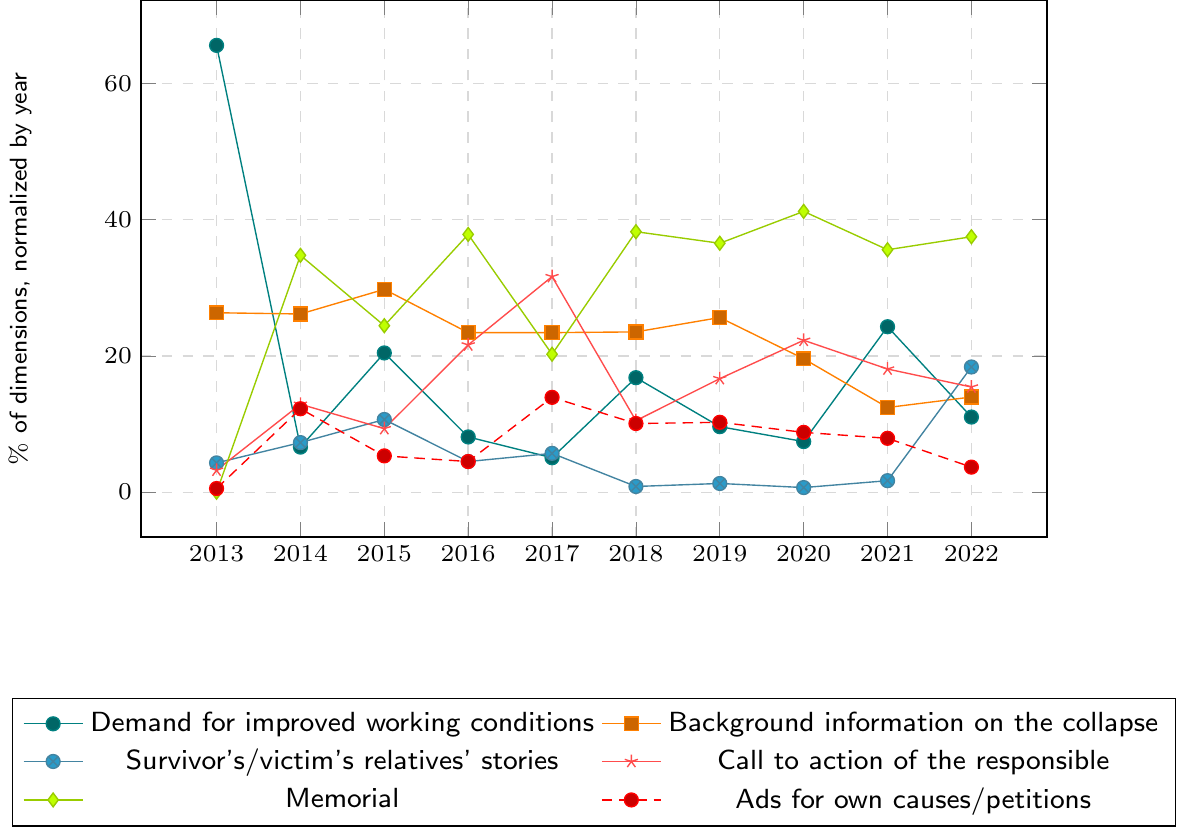}
	\end{center}
	\caption{Relative occurrences of the dimensions identified by the SKAD by year.}\label{fig:agg_codes_by_year}
\end{figure}

\begin{figure}
	\begin{center}
		\includegraphics[width=.99\textwidth]{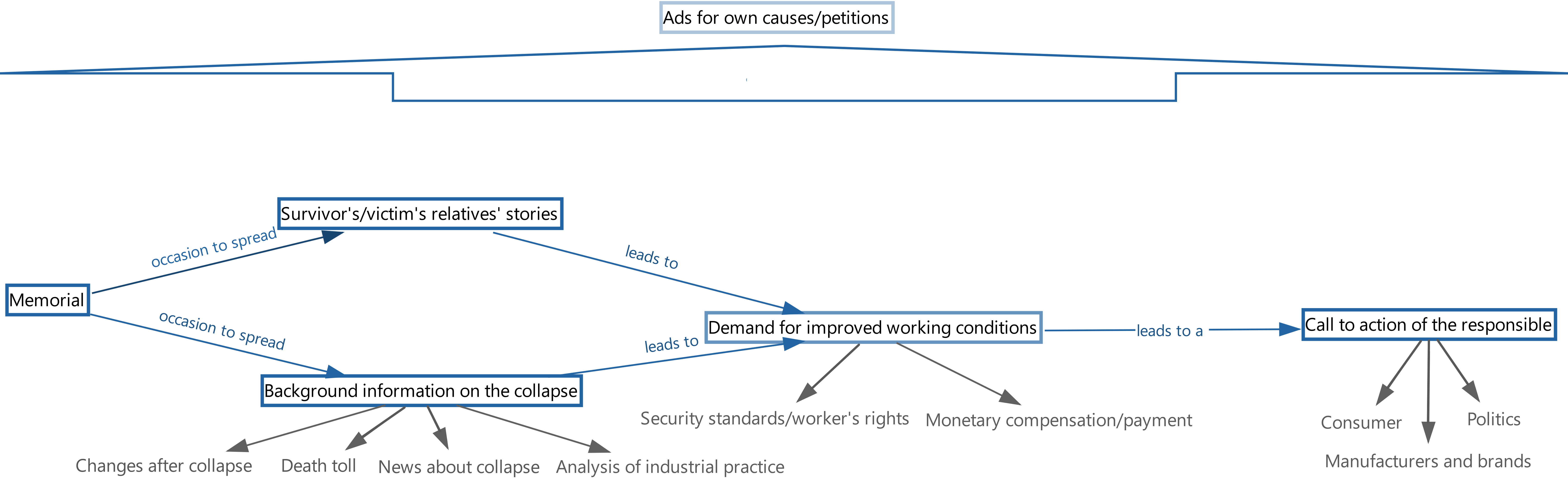}
	\end{center}
	\caption{Schematic illustration of contentual relations between the dimensions and codes as a result of the coding process.
		Codes are marked gray, the five dark blue dimensions induce the light blue dimension \texttt{Ads for own causes}.}\label{fig:code_network}
\end{figure}

The first result of the qualitative analysis by the sociology of knowledge approach to discourse (SKAD) described in \Cref{sec:methods_content} is a set of codes obtained in the coding process of all tweets by the ten most central users per year.
\Cref{tab:code_overview} lists the identified dimensions alongside codes obtained in the coding process as well as a short description and one example tweet.
Furthermore, \Cref{fig:agg_codes_by_year} shows the relative prevalence of the six dimensions per year.
Finally, \Cref{fig:code_network} depicts the contentual relationships between the dimensions and codes identified in the coding step of the SKAD.
It schematically summarizes the communicative pattern that the anniversary is taken as an opportunity to first raise awareness with memorials and then provide background information and describe working conditions, thus justify the demand for action.

\begin{table}
	\scriptsize
	\begin{tabular}{p{3cm}p{6cm}p{6cm}}
		\hline\hline
		\textbf{Dimensions}/Codes & Description & Example tweet\\\hline\hline
		\textbf{Memorial} & Remembering the anniversary of the collapse & \textit{In the runup to the anniversary of the Rana Plaza collapse, we commemorate those we lost, we pay tribute to those w... https://t.co/AM5ir9l2k2}, OhSoEthical, 2022\\\hline\hline
		\textbf{Survivor's/victim's relatives' stories} & Reports, pictures, and stories of the living conditions of the victims and their families & \textit{"RANA PLAZA 9 YEARS ON: "I wish I had not survived Rana Plaza"}
		\textit{Rana Plaza exposed the sheer brutality of the fash... https://t.co/ixW1O4sMXw}, nijjormanush, 2022\\\hline\hline
		\textbf{Background information on the collapse} & Details about the collapse of the Rana Plaza factory itself &\\\hline
		Changes after collapse & (Photo) Reports on what has and hasn't changed over the years & \textit{Rana Plaza: three years on, garment workers still exploited \#RanaPlaza \#LoveFashionHateSweatshops https://t.co/7C2Kj2DmW0}, WarOnWant, 2016\\\hline
		Death toll & Count of the victims of the collapse & \textit{More than 700 people didn't survive the Bangladesh collapse. Tell brands to sign the Safety Agreement and stop this. http://t.co/mh8OHZi6Pr}, cleanclothes, 2013\\\hline
		News about collapse & Information on what has happened before, during, and after the collapse & \textit{In the days that preceded the Rana Plaza tragedy, cracks appeared in the building walls and workers expressed fears... https://t.co/UGSTfracKN}, Fash\_Rev, 2020\\\hline
		Analysis of industrial practice & Bringing the Rana Plaza collapse into a bigger picture, e.g., global interconnection of the fashion industry & \textit{``Sustainability isn't just about the environment--it’s much more of a social issue.'' https://t.co/rrIPeG3bgz}, voguemagazine, 2021\\\hline\hline
		\textbf{Demand for improved working conditions} & Demand for improved working conditions in Bangladesh & \\\hline
		Security standards/ worker's rights & Demand for improved security standards, e.g., fire safety or the Accord agreement and worker's rights, e.g., worker's unions & \textit{Support workers in the garment industry on \#MayDay Tell brands to make sure their work place is safe! http://t.co/mh8OHZi6Pr}, cleanclothes, 2013\\\hline
		Monetary compensation/payment & Demand for higher wages and compensations for the Rana Plaza victims -- mentions of \#payup or the Rana Plaza fund & \textit{Many survivors of \#Bangladesh factory collapse say they face serious economic hardship | http://t.co/j8zSYEZs2A http://t.co/ddhz3wyJEr}, AJEnglish, 2014\\\hline\hline
		\textbf{Call to action of the responsible} & Demand for change of the involved actor groups &\\\hline
		Consumer & Demand for change in fashion consumption & \textit{Though consumers were not the cause of the problem, they can be part of the solution @Carrysomers in @glammonitor http://t.co/FXZSjhTNY5}, Fash\_Rev, 2015\\\hline
		Manufacturers and brands & Demand for change of industrial practices & \textit{Two years from \#RanaPlaza disaster: brands still owe US\$ 6 million to the 30 mn compensation fund @IndustriALL\_GU http://t.co/o5w8PPYjwy}, IndustriALL\_GU, 2015\\\hline
		Politics & Demand for change of policies & \textit{3 years on from Rana Plaza, EU co's still not obligated to prevent human rights abuses in supply chain https://t.co/dxQ3UsRJeh \#FashRev}, Fash\_Rev, 2016\\\hline\hline
		\textbf{Ads for own causes/petitions} & Simple advertisement of own projects & \textit{Trace My Fashion project launched by @Fash\_Rev\_BD in Bangladesh is featured in today's @guardian http://t.co/f6tCSyz6fF \#whomademyclothes}, Fash\_Rev, 2015\\\hline\hline
	\end{tabular}
	\caption{List of dimensions and codes identified in the coding process of the SKAD.
		Dimensions are marked in bold font.
		The three dimensions \texttt{Demand for improved working conditions}, \texttt{Background information on the collapse}, and \texttt{Call to action of the responsible} have corresponding codes.}\label{tab:code_overview}
\end{table}

A detailed analysis of the individual years reveals a strong heterogeneity across the years.
In $2013$, we detect three main codes \texttt{Security standards/worker's rights}, \texttt{Death toll}, and \texttt{News about collapse} belonging to the two dimensions \texttt{Demand for improved working conditions} and \texttt{Background information on the collapse}.
The codes \texttt{Security standards/worker's rights} and \texttt{News about collapse} are the most prevalent topics as both cover over $50\%$ of their respective code occurrences among all years and are often published by Asian actors.

\begin{sloppypar}
At the first anniversary in the year $2014$, we observe a peak occurrence of eight codes and in comparison to other years a wide range of topics being discussed intensively.
Especially compared to the news feed style of the discourse in $2013$ we find more in-depth and complex content such as background analyses, individual stories, or reports about what has changed since the collapse.
\Cref{fig:agg_codes_by_year} shows that while the dimensions \texttt{Demand for improved working conditions} and \texttt{Background information on the collapse} peak in $2013$, all other dimensions start to emerge in $2014$.
\texttt{Analysis of industrial practice} as well as \texttt{Survivor's/victim's relatives' stories} are mainly published in $2014$ and $2015$ indicating that in $2013$ the news were too recent for such discussions even two weeks after the incident\footnote{We considered a time horizon of one week before and approximately two weeks after the anniversary date of each year, cf.\ \Cref{sec:data}.}.
Last but not least, the dimension \texttt{Memorial} indicating the emergence of commemorative practices establishes itself in $2014$ as the most frequent code in all and the most frequent dimension in almost all years following $2014$.
\end{sloppypar}

The years $2015$ to $2022$ are characterized by the repetition of the interpretative dimensions established in $2014$ with some variations of major topics emerging for individual years.
In $2015$, for example, we observe the peak of the code \texttt{Monetary compensation/payment} as the ``Rana Plaza Fund'' is publicly raising money in order to pay compensation to the victims.
In total, the fund raised $30$ mio.\ USD for Rana Plaza victims with $1.1$ mio.\ USD donated by the brand ``benetton'' \cite{ranaplazaarrangement,hoskins2015after}.
This brand's role is special in the sense that it took an active role in the Twitter discourse: $2015$, ``benetton'' is the only brand appearing as a top ten central user across all years.

In $2016$, the total number of tweets decreases and besides \texttt{Memorial} the main topics are the responsibility of \texttt{Manufacturers and brands} as well as the \texttt{Analysis of industrial practice}.
In $2017$, the total number of tweets slightly increases again and \texttt{Consumer} orientation reaches its peak -- caused by the retweet of the Guardian article ``Sustainable style: will Gen Z help the fashion industry clean up its act?'' by ``EmmaWatson'' on April $27$th $2017$.
This retweet not only leads to the distinct peak in tweet volume that can be observed in \Cref{fig:daily_tweet_volume}; it also establishes ``EmmaWatson'' as the most central user in year $2017$ and sets the agenda of topics discussed by other central users.

In the context of the COVID-19 pandemic and the withdrawal of the Bangladesh Accord agreement \cite{salminen2018accord}, the years $2020$ and $2021$ witness another rise of the code \texttt{Analysis of industrial practice}.
In $2022$, \texttt{Survivor's/victim's relatives' stories} suddenly rise to their peak -- a thematic emphasis that can be traced back to the activists behind the user ``OhSoEthical''.
Albeit frequent repetition of the code by that user, the community detection result in Figure~\ref{fig:centralities_communities} reveals that only a relatively small community is reached by these tweets.

The codes \texttt{Memorial} and \texttt{Death toll} are identified as an incentive to spread further information on the background of the incident.
Quantitatively, this trigger constitutes up to $40\%$ of all codes, making it an important starting point in the Twitter discourse.
The survivor stories and background information lead to a constant call for action to change the described circumstances in the fashion industry.

\section{Discussion}\label{sec:discussion}

\begin{table}
	\setlength{\tabcolsep}{1pt}
	\begin{tabular}{llll}
		&\parbox[t]{0mm}{\rotatebox[origin=l]{90}{Reporting}}&\parbox[t]{0mm}{\rotatebox[origin=l]{90}{Reprocessing}}&\parbox[t]{0mm}{\rotatebox[origin=l]{90}{Commemoration}}\\
		\begin{tabular}{l}
		pase mas\\
		ropalimpia (13)
		\end{tabular}&\deepskybluebar{.5}&\bluebar{0}&\greenbar{0}\\
		\begin{tabular}{l}
		garment\\
		industry (19)
		\end{tabular}&\deepskybluebar{.5}&\bluebar{0.333}&\greenbar{.292}\\
		tell brands (16)&\deepskybluebar{.5}&\bluebar{0}&\greenbar{0}\\
		ropalimpia (100)&\deepskybluebar{.5}&\bluebar{0}&\greenbar{0.0006}\\
		brands (95)&\deepskybluebar{.5}&\bluebar{.1667}&\greenbar{.246}\\
		acuerdo (92)&\deepskybluebar{.5}&\bluebar{0}&\greenbar{.002}\\
		seguridad (90)&\deepskybluebar{.5}&\bluebar{0}&\greenbar{.0007}\\
	\end{tabular}
	\hspace{2pt}
	\begin{tabular}{llll}
		&\parbox[t]{0mm}{\rotatebox[origin=l]{90}{Reporting}}&\parbox[t]{0mm}{\rotatebox[origin=l]{90}{Reprocessing}}&\parbox[t]{0mm}{\rotatebox[origin=l]{90}{Commemoration}}\\
		\begin{tabular}{l}
		fashion revolution\\
		day (27)
		\end{tabular}&\deepskybluebar{0}&\bluebar{0.5}&\greenbar{.026}\\
		lost their lives (14)&\deepskybluebar{0}&\bluebar{.5}&\greenbar{.375}\\
		fashion revolution (46)&\deepskybluebar{0}&\bluebar{0.5}&\greenbar{.074}\\
		factory collapse (46)&\deepskybluebar{.1}&\bluebar{.5}&\greenbar{.212}\\
		your clothes (19)&\deepskybluebar{0}&\bluebar{.5}&\greenbar{.056}\\
		building collapse (14)&\deepskybluebar{0}&\bluebar{.5}&\greenbar{.229}\\
		collapse (152)&\deepskybluebar{.142}&\bluebar{.5}&\greenbar{.372}\\
		garment (119)&\deepskybluebar{.139}&\bluebar{.5}&\greenbar{.33}\\
		anniversary (113)&\deepskybluebar{0}&\bluebar{.5}&\greenbar{.245}\\
		fashion (106)&\deepskybluebar{0}&\bluebar{.5}&\greenbar{.106}\\
		after (103)&\deepskybluebar{.048}&\bluebar{.5}&\greenbar{.238}\\
		factory (92)&\deepskybluebar{.083}&\bluebar{.5}&\greenbar{.247}\\
		today (92)&\deepskybluebar{0}&\bluebar{.5}&\greenbar{.276}\\
		disaster (90)&\deepskybluebar{0}&\bluebar{.5}&\greenbar{.193}\\
	\end{tabular}
	\hspace{2pt}
	\begin{tabular}{llll}
		&\parbox[t]{0mm}{\rotatebox[origin=l]{90}{Reporting}}&\parbox[t]{0mm}{\rotatebox[origin=l]{90}{Reprocessing}}&\parbox[t]{0mm}{\rotatebox[origin=l]{90}{Commemoration}}\\
		\begin{tabular}{l}
		garment\\
		workers (65)
		\end{tabular}&\deepskybluebar{0}&\bluebar{.262}&\greenbar{.5}\\
		workers (163)&\deepskybluebar{.11}&\bluebar{.386}&\greenbar{.5}\\
	\end{tabular}
	\vspace{10pt}
	\caption{Word (combination) frequencies in all tweets by the ten most central users across the ten years.
	Numbers in round bracket indicate the absolute number of word occurrences in all years.
	The bar charts denote yearly word frequencies, i.e., word occurrences in the commemoration phase are averaged over the years $2015$ to $2022$.
	Furthermore, each word (combination)'s bar charts are normalized by the respective maximum number of occurrences across the three phases.}\label{tab:word_frequency_bars}
\end{table}

The results presented in the previous section suggest that the international Twitter discourse on the Rana Plaza collapse is composed of three phases: reporting, reprocessing, and commemoration.
These phases follow the temporal unfolding of the discourse and can be identified from both a contentual qualitative as well as a structural quantitative viewpoint.

The first phase, reporting, starts with the collapse date in $2013$ and is mainly formed by Asian and Western news agencies informing the public about the accident and its extent, e.g., the death toll.
Furthermore, a comparatively small but active community of NGOs explicitly addresses manufacturers and brands demanding better working conditions of textile workers in Bangladesh.
This is reflected by the demand for an ``acuerdo de seguridad'' (engl.: safety agreement) that a group of Spanish activists directs towards brands, cf.~\Cref{tab:word_frequency_bars}.
	Additionally, this is underpinned by the three main codes \texttt{Security standards/worker's rights}, \texttt{Death toll}, and \texttt{News about collapse} obtained in the coding phase of the SKAD as well as by the prevalence of news agencies as most and NGOs as second most frequent actor group in the centrality analysis.
The communities emerging around central users show remarkably little interaction indicating the formation of communication bubbles around different news agencies and NGOs.
In terms of the involved user's nationalities, a relatively large portion of users with a small physical proximity to the Rana Plaza factory such as Bangladesh or Indonesia shape the $2013$ discourse.

The second phase, reprocessing, takes place around the first anniversary of the collapse in $2014$.
The overall tweet volume, the number of involved users, as well as the structural and contentual complexity of the discourse all reach their peaks.
The active discussion of various social, political, and economical aspects of the collapse reflects the societal negotiation process establishing a collective interpretative framework of the event.
This variety of topics is, for instance, reflected in the word frequency analysis in \Cref{tab:word_frequency_bars} showing that the most important word (combination) frequencies reach their peak in the reprocessing phase and prevail in the consecutive commemoration phase.
In particular, the first anniversary witnesses the rise of the dimension \texttt{Memorial} as most frequent dimension, which establishes a pattern repeated in subsequent years.
These simple commemorations serve as a ramp for other dimensions;
\texttt{Survivor's/victim's relatives' stories} and \texttt{Background information on the collapse} are taken as a peg to address responsibilities and to demand consequences.
Most actor groups such as NGOs or writers accompany their contribution by \texttt{Ads for own causes/petitions}.

Regarding speaker positions, the centrality analysis reveals that in $2014$ mainly NGOs seize control of the Twitter discourse by lobbying for consequences and changes in the fast fashion industry.
Furthermore, the active dynamics of the $2014$ discussion are reflected in the high degree of interconnectivity across different (relatively large) communities and actor groups.
The prominence of the widely used hashtag \#insideout is conjectured to additionally promote interactions between heterogeneous user groups.
While the tweet volume of countries with close spatial proximity to the Rana Plaza factory decreases, $2014$ witnesses a distinct rise of discourse participation by Western countries -- suggesting that the interpretative momentum, i.e., the discursive power is over-proportionally seized by Western actors.

The third phase, commemoration, spans all subsequently investigated years from $2015$ to $2022$.
It is mainly shaped by the repetition of the interpretative patterns negotiated in the reprocessing phase in $2014$ and is complemented by uprising discussions on contemporary topics, which are discussed in detail in \Cref{sec:results_content}.
While, overall, we observe high fluctuations in the user base of the Rana Plaza remembrance community over the years, cf.\ \Cref{sec:results_multiplex}, the active NGO user group that formed in $2014$ continues to dominate the discourse and inter-community connectivity tends to take place between NGO communities.
More specifically, we detect few interconnected key actors such as ``Fash\_Rev'' and ``cleanclothes'' re-appearing in the top ten most central users almost every year.
While we focus on the Twitter discourse in this work, ``Fash\_Rev'' (Fashion Revolution) also organizes offline activities such as the yearly Fashion Revolution week\footnote{cf.\ e.g., \url{https://www.fashionrevolution.org/frw-2022/}} around the Rana Plaza anniversary, which in turn influences the online discourse by stimulating discussions on these offline activities.

\section{Conclusion}\label{sec:conclusion}

In this work, we studied the global Twitter discourse on the Rana Plaza factory collapse by a mixture of established quantitative network-theoretic and qualitative discourse-theoretic methods.
We analyzed structural network properties and the temporal evolution of interpretative patterns of the collapse over the years $2013$ to $2022$ and found a division of the discourse into three phases: reporting, reprocessing, and commemoration.
Our analysis reveals, which actors and communities worldwide interact, build, and rebuild the Rana Plaza discourse on Twitter every year.

Further analysis could focus on how our findings relate to existing theories on key events \cite{brosius1995prototyping} and digital memory culture \cite{garde2009save,pentzold20222} or how the Twitter discourse is embedded into the general discourse across various communication channels.
Additionally, further structural and dynamical multiplex network properties could be examined.
For instance, it would be interesting to study centralities of the full ten year time horizon, which may be possible by a multilayer modeling approach taking a second aspect representing time into account.
Studying the applicability of matrix function-based centralities to this type of network architecture would be an interesting road for future research.
Furthermore, preliminary investigations of the years $2013$ and $2014$ indicate a strong multilayer core--periphery structure \cite{bergermann2023nonlinear}.

\section*{Acknowledgements}
We thank the anonymous	referees for their helpful comments.

\section*{Abbreviations}
SKAD: Sociology of knowledge approach to discourse;
NGO: Non-governmental organization;
API: Application programming interface;
TC: Total communicability;
KC: Katz centrality.

\section*{Availability of data and materials}
The networks generated in this study are publicly available in Matlab and python format under 
\url{https://github.com/KBergermann/Twitter-Rana-Plaza}.
The same repository contains publicly available codes to reproduce all results obtained by the computational methods described in \Cref{sec:methods_multiplex_centralities,sec:methods_multiplex_communities}.


\begin{thebibliography}{10}
	
	\bibitem{bair2020political}
	{\sc J.~Bair, M.~Anner, and J.~Blasi}, {\em The political economy of private
		and public regulation in post-{R}ana {P}laza {B}angladesh}, ILR Review, 73
	(2020), pp.~969--994.
	
	\bibitem{barua2017workplace}
	{\sc U.~Barua and M.~A. Ansary}, {\em Workplace safety in {B}angladesh
		ready-made garment sector: 3 years after the {R}ana {P}laza collapse},
	International Journal of Occupational Safety and Ergonomics, 23 (2017),
	pp.~578--583.
	
	\bibitem{benzi2020matrix}
	{\sc M.~Benzi and P.~Boito}, {\em Matrix functions in network analysis},
	GAMM-Mitteilungen, 43 (2020), p.~e202000012.
	
	\bibitem{benzi2013total}
	{\sc M.~Benzi and C.~Klymko}, {\em Total communicability as a centrality
		measure}, Journal of Complex Networks, 1 (2013), pp.~124--149.
	
	\bibitem{benzi2015limiting}
	\leavevmode\vrule height 2pt depth -1.6pt width 23pt, {\em On the limiting
		behavior of parameter-dependent network centrality measures}, SIAM Journal on
	Matrix Analysis and Applications, 36 (2015), pp.~686--706.
	
	\bibitem{berger1967social}
	{\sc P.~L. Berger and T.~Luckmann}, {\em The Social Construction of Reality: A
		Treatise in the Sociology of Knowledge}, Anchor, USA, 1967.
	
	\bibitem{bergermannMMFC}
	{\sc K.~Bergermann}, {\em Code release:
		Multiplex-matrix-function-centralities}, Available at
	\url{https://github.com/KBergermann/Multiplex-matrix-function-centralities},
	(2021).
	
	\bibitem{bergermannUrban}
	\leavevmode\vrule height 2pt depth -1.6pt width 23pt, {\em Code release:
		Urban-multiplex-networks}, Available at
	\url{https://github.com/KBergermann/Urban-multiplex-networks},  (2021).
	
	\bibitem{bergermannTwitter}
	\leavevmode\vrule height 2pt depth -1.6pt width 23pt, {\em Code release:
		Twitter-{R}ana-{P}laza}, Available at
	\url{https://github.com/KBergermann/Twitter-Rana-Plaza},  (2023).
	
	\bibitem{bergermann2021orientations}
	{\sc K.~Bergermann and M.~Stoll}, {\em Orientations and matrix function-based
		centralities in multiplex network analysis of urban public transport},
	Applied Network Science, 6 (2021), pp.~1--33.
	
	\bibitem{bergermann2022fast}
	\leavevmode\vrule height 2pt depth -1.6pt width 23pt, {\em Fast computation of
		matrix function-based centrality measures for layer-coupled multiplex
		networks}, Physical Review E, 105 (2022), p.~034305.
	
	\bibitem{bergermann2023nonlinear}
	{\sc K.~Bergermann, M.~Stoll, and F.~Tudisco}, {\em A nonlinear spectral
		core-periphery detection method for multiplex networks}, In preparation,
	(2023).
	
	\bibitem{blondel2008fast}
	{\sc V.~D. Blondel, J.-L. Guillaume, R.~Lambiotte, and E.~Lefebvre}, {\em Fast
		unfolding of communities in large networks}, Journal of Statistical
	Mechanics: Theory and Experiment, 2008 (2008), p.~P10008.
	
	\bibitem{boccaletti2014structure}
	{\sc S.~Boccaletti, G.~Bianconi, R.~Criado, C.~I. Del~Genio,
		J.~G{\'o}mez-Gardenes, M.~Romance, I.~Sendina-Nadal, Z.~Wang, and M.~Zanin},
	{\em The structure and dynamics of multilayer networks}, Physics Reports, 544
	(2014), pp.~1--122.
	
	\bibitem{bonacich1987power}
	{\sc P.~Bonacich}, {\em Power and centrality: {A} family of measures}, American
	Journal of Sociology, 92 (1987), pp.~1170--1182.
	
	\bibitem{borgatti2018analyzing}
	{\sc S.~P. Borgatti, M.~G. Everett, and J.~C. Johnson}, {\em Analyzing Social
		Networks}, Sage, USA, 2018.
	
	\bibitem{brandes2013social}
	{\sc U.~Brandes, L.~C. Freeman, and D.~Wagner}, {\em Social networks}, in
	Handbook of Graph Drawing and Visualization, Chapman \& Hall, UK, 2013,
	pp.~805--839.
	
	\bibitem{brin1998anatomy}
	{\sc S.~Brin and L.~Page}, {\em The anatomy of a large-scale hypertextual web
		search engine}, Computer networks and ISDN systems, 30 (1998), pp.~107--117.
	
	\bibitem{brosius1995prototyping}
	{\sc H.-B. Brosius and P.~Eps}, {\em Prototyping through key events: News
		selection in the case of violence against aliens and asylum seekers in
		germany}, European Journal of Communication, 10 (1995), pp.~391--412.
	
	\bibitem{ch2015local}
	{\sc E.~Ch'ng}, {\em Local interactions and the emergence of a twitter
		small-world network}, arXiv preprint \url{https://arxiv.org/abs/1508.03594},
	(2015).
	
	\bibitem{chowdhury2017rana}
	{\sc R.~Chowdhury}, {\em The {R}ana {P}laza disaster and the complicit behavior
		of elite {NGO}s}, Organization, 24 (2017), pp.~938--949.
	
	\bibitem{dickison2016multilayer}
	{\sc M.~E. Dickison, M.~Magnani, and L.~Rossi}, {\em Multilayer Social
		Networks}, Cambridge University Press, UK, 2016.
	
	\bibitem{estrada2010network}
	{\sc E.~Estrada and D.~J. Higham}, {\em Network properties revealed through
		matrix functions}, SIAM {R}eview, 52 (2010), pp.~696--714.
	
	\bibitem{estrada2005subgraph}
	{\sc E.~Estrada and J.~A. Rodriguez-Velazquez}, {\em Subgraph centrality in
		complex networks}, Physical Review E, 71 (2005), p.~056103.
	
	\bibitem{fortunato2010community}
	{\sc S.~Fortunato}, {\em Community detection in graphs}, Physics Reports, 486
	(2010), pp.~75--174.
	
	\bibitem{foucault1970archaeology}
	{\sc M.~Foucault}, {\em The archaeology of knowledge}, Social Science
	Information, 9 (1970), pp.~175--185.
	
	\bibitem{foucault2005order}
	\leavevmode\vrule height 2pt depth -1.6pt width 23pt, {\em The Order of
		Things}, Routledge, UK, 2005.
	
	\bibitem{foucault1997polemics}
	{\sc M.~Foucault, P.~Rabinow, and R.~Hurley}, {\em Polemics, {P}olitics, and
		{P}roblematizations: {A}n interview with {M}ichel {F}oucault. {I}: {E}thics,
		{S}ubjectivity and {T}ruth}, 1997.
	
	\bibitem{freeman1977set}
	{\sc L.~C. Freeman}, {\em A set of measures of centrality based on
		betweenness}, Sociometry, 40 (1977), pp.~35--41.
	
	\bibitem{freeman1978centrality}
	\leavevmode\vrule height 2pt depth -1.6pt width 23pt, {\em Centrality in social
		networks conceptual clarification}, Social Networks, 1 (1978), pp.~215--239.
	
	\bibitem{garde2009save}
	{\sc J.~Garde-Hansen, A.~Hoskins, and A.~Reading}, {\em Save as... digital
		memories}, Springer, Germany, 2009.
	
	\bibitem{girvan2002community}
	{\sc M.~Girvan and M.~E. Newman}, {\em Community structure in social and
		biological networks}, Proceedings of the National Academy of Sciences, 99
	(2002), pp.~7821--7826.
	
	\bibitem{glaser1967discovery}
	{\sc B.~G. Glaser and A.~L. Strauss}, {\em The Discovery of Grounded Theory:
		Strategies for Qualitative Research}, Routledge, UK, 1967.
	
	\bibitem{hanteer2019innovative}
	{\sc O.~Hanteer and L.~Rossi}, {\em An innovative way to model {T}witter
		topic-driven interactions using multiplex networks}, Frontiers in big Data,
	(2019), p.~9.
	
	\bibitem{higham2008functions}
	{\sc N.~J. Higham}, {\em Functions of {M}atrices: {T}heory and {C}omputation},
	SIAM, USA, 2008.
	
	\bibitem{hoskins2015after}
	{\sc T.~Hoskins}, {\em After two years, the {R}ana {P}laza fund finally reaches
		its \$30m target}, The Guardian,  (2015).
	
	\bibitem{JeubGenLouvain}
	{\sc L.~G.~S. Jeub, M.~Bazzi, I.~S. Jutla, and P.~J. Mucha}, {\em Code release:
		A generalized {L}ouvain method for community detection implemented in
		{MATLAB}}, Available at \url{https://github.com/GenLouvain/GenLouvain},
	(2011-2019).
	
	\bibitem{katz1953new}
	{\sc L.~Katz}, {\em A new status index derived from sociometric analysis},
	Psychometrika, 18 (1953), pp.~39--43.
	
	\bibitem{keller2011sociology}
	{\sc R.~Keller}, {\em The sociology of knowledge approach to discourse
		({SKAD})}, Human Studies, 34 (2011), pp.~43--65.
	
	\bibitem{kivela2014multilayer}
	{\sc M.~Kivel{\"a}, A.~Arenas, M.~Barth{\'e}lemy, J.~P. Gleeson, Y.~Moreno, and
		M.~A. Porter}, {\em Multilayer networks}, Journal of Complex Networks, 2
	(2014), pp.~203--271.
	
	\bibitem{kleinberg1999authoritative}
	{\sc J.~M. Kleinberg}, {\em Authoritative sources in a hyperlinked
		environment}, Journal of the ACM (JACM), 46 (1999), pp.~604--632.
	
	\bibitem{knoblauch2019communicative}
	{\sc H.~Knoblauch}, {\em The Communicative Construction of Reality}, Routledge,
	UK, 2019.
	
	\bibitem{martin2014localization}
	{\sc T.~Martin, X.~Zhang, and M.~E. Newman}, {\em Localization and centrality
		in networks}, Physical Review E, 90 (2014), p.~052808.
	
	\bibitem{mucha2010community}
	{\sc P.~J. Mucha, T.~Richardson, K.~Macon, M.~A. Porter, and J.-P. Onnela},
	{\em Community structure in time-dependent, multiscale, and multiplex
		networks}, Science, 328 (2010), pp.~876--878.
	
	\bibitem{newman2003structure}
	{\sc M.~E. Newman}, {\em The structure and function of complex networks}, SIAM
	Review, 45 (2003), pp.~167--256.
	
	\bibitem{newman2006modularity}
	\leavevmode\vrule height 2pt depth -1.6pt width 23pt, {\em Modularity and
		community structure in networks}, Proceedings of the National Academy of
	Sciences, 103 (2006), pp.~8577--8582.
	
	\bibitem{omodei2015characterizing}
	{\sc E.~Omodei, M.~De~Domenico, and A.~Arenas}, {\em Characterizing
		interactions in online social networks during exceptional events}, Frontiers
	in Physics, 3 (2015), p.~59.
	
	\bibitem{page1999pagerank}
	{\sc L.~Page, S.~Brin, R.~Motwani, and T.~Winograd}, {\em The {P}age{R}ank
		citation ranking: {B}ringing order to the web}, tech. rep., Stanford InfoLab,
	1999.
	
	\bibitem{papadopoulos2012community}
	{\sc S.~Papadopoulos, Y.~Kompatsiaris, A.~Vakali, and P.~Spyridonos}, {\em
		Community detection in social media}, Data Mining and Knowledge Discovery, 24
	(2012), pp.~515--554.
	
	\bibitem{pentzold20222}
	{\sc C.~Pentzold, C.~Lohmeier, and T.~Birkner}, {\em 2 {K}ommunikatives
		{E}rinnern}, in Handbuch kommunikationswissenschaftliche
	Erinnerungsforschung, De Gruyter, Germany, 2022, pp.~47--70.
	
	\bibitem{pina2016towards}
	{\sc C.~Pi{\~n}a-Garc{\'\i}a, C.~Gershenson, and J.~M. Siqueiros-Garc{\'\i}a},
	{\em Towards a standard sampling methodology on online social networks:
		collecting global trends on twitter}, Applied Network Science, 1 (2016),
	pp.~1--19.
	
	\bibitem{rahman2020multi}
	{\sc S.~Rahman and K.~M. Rahman}, {\em Multi-actor initiatives after {R}ana
		{P}laza: {F}actory managers’ views}, Development and change, 51 (2020),
	pp.~1331--1359.
	
	\bibitem{reguero2023journalism}
	{\sc I.~Reguero-Sanz, P.~Berd{\'o}n-Prieto, and J.~Herrero-Izquierdo}, {\em
		Journalism in democracy: A discourse analysis of {T}witter posts on the
		ferrerasgate scandal}, Media and Communication, 11 (2023).
	
	\bibitem{rehman2020identification}
	{\sc A.~U. Rehman, A.~Jiang, A.~Rehman, A.~Paul, M.~T. Sadiq, et~al.}, {\em
		Identification and role of opinion leaders in information diffusion for
		online discussion network}, Journal of Ambient Intelligence and Humanized
	Computing,  (2020), pp.~1--13.
	
	\bibitem{reinecke2015after}
	{\sc J.~Reinecke and J.~Donaghey}, {\em After {R}ana {P}laza: {B}uilding
		coalitional power for labour rights between unions and (consumption-based)
		social movement organisations}, Organization, 22 (2015), pp.~720--740.
	
	\bibitem{riquelme2016measuring}
	{\sc F.~Riquelme and P.~Gonz{\'a}lez-Cantergiani}, {\em Measuring user
		influence on {T}witter: {A} survey}, Information Processing \& Management, 52
	(2016), pp.~949--975.
	
	\bibitem{sadri2020exploring}
	{\sc A.~M. Sadri, S.~Hasan, S.~V. Ukkusuri, and M.~Cebrian}, {\em Exploring
		network properties of social media interactions and activities during
		{H}urricane {S}andy}, Transportation Research Interdisciplinary Perspectives,
	6 (2020), p.~100143.
	
	\bibitem{salminen2018accord}
	{\sc J.~Salminen}, {\em The accord on fire and building safety in {B}angladesh:
		A new paradigm for limiting buyers’ liability in global supply chains?},
	The American Journal of Comparative Law, 66 (2018), pp.~411--451.
	
	\bibitem{scott2012social}
	{\sc J.~Scott}, {\em What is social network analysis?}, Bloomsbury Academic,
	UK, 2012.
	
	\bibitem{shea2022david}
	{\sc C.~S. Shea, Y.~Jiang, and W.~L. Leung}, {\em David vs. {G}oliath:
		Transnational grassroots outreach and empirical evidence from the
		\#{H}ong{K}ong{P}rotests {T}witter network}, Review of Communication, 22
	(2022), pp.~193--212.
	
	\bibitem{siddiqui2016human}
	{\sc J.~Siddiqui and S.~Uddin}, {\em Human rights disasters, corporate
		accountability and the state: {L}essons learned from {R}ana {P}laza},
	Accounting, Auditing \& Accountability Journal,  (2016).
	
	\bibitem{taylor2017eigenvector}
	{\sc D.~Taylor, S.~A. Myers, A.~Clauset, M.~A. Porter, and P.~J. Mucha}, {\em
		Eigenvector-based centrality measures for temporal networks}, Multiscale
	Modeling \& Simulation, 15 (2017), pp.~537--574.
	
	\bibitem{ranaplazaarrangement}
	{\sc The{\ }{R}ana{\ }{P}laza{\ }Arrangement},
	\url{https://ranaplaza-arrangement.org/},  ((accessed on March 3rd, 2023)).
	
	\bibitem{von2007tutorial}
	{\sc U.~Von~Luxburg}, {\em A tutorial on spectral clustering}, Statistics and
	computing, 17 (2007), pp.~395--416.
	
	\bibitem{waters2011tweet}
	{\sc R.~D. Waters and J.~Y. Jamal}, {\em Tweet, tweet, tweet: A content
		analysis of nonprofit organizations’ {T}witter updates}, Public Relations
	Review, 37 (2011), pp.~321--324.
	
	\bibitem{watts1998collective}
	{\sc D.~J. Watts and S.~H. Strogatz}, {\em Collective dynamics of
		‘small-world’ networks}, Nature, 393 (1998), pp.~440--442.
	
	\bibitem{wicke2021covid}
	{\sc P.~Wicke and M.~M. Bolognesi}, {\em Covid-19 discourse on {T}witter: How
		the topics, sentiments, subjectivity, and figurative frames changed over
		time}, Frontiers in Communication, 6 (2021), p.~651997.
	
	\bibitem{wiggins2022nothing}
	{\sc B.~Wiggins}, {\em ‘{N}othing {C}an {S}top {W}hat’s {C}oming’: {A}n
		analysis of the conspiracy theory discourse on 4chan’s/{P}ol board},
	Discourse \& Society,  (2022).
	
	\bibitem{wu2018emotional}
	{\sc D.~D. Wu and C.~Li}, {\em Emotional branding on social media: A
		cross-cultural discourse analysis of global brands on {T}witter and {W}eibo},
	Intercultural communication in Asia: Education, language and values,  (2018),
	pp.~225--240.
	
	\bibitem{yaqub2017analysis}
	{\sc U.~Yaqub, S.~A. Chun, V.~Atluri, and J.~Vaidya}, {\em Analysis of
		political discourse on {T}witter in the context of the 2016 {US} presidential
		elections}, Government Information Quarterly, 34 (2017), pp.~613--626.
	
\end{thebibliography}
\end{document}